\documentclass[10pt]{iopart}

\usepackage{iopams}
\usepackage{bm}
\usepackage{bbold}

\usepackage[percent]{overpic}

\usepackage{comment}

\usepackage {ulem}

\usepackage {graphicx}
\usepackage{color}

\def\>{\right\rangle}
\def\<{\left\langle}
\def\be{\begin{equation}}
\def\ee{\end{equation}}
\def\ba{\begin{array}{lll}}
\def\ea{\end{array}}

\def\beq{\begin{eqnarray}}
\def\eeq{\end{eqnarray}}

\newcommand{\CGF}{ G_\nu}

\newcommand{\average}[1]{\langle #1 \rangle}
\newcommand{\Trace}{{\rm Tr}}

\newcommand{\D}{\mathcal{D}}

\newcommand{\matrixel}[3]{{\mathinner{\langle{#1}| {#2} | {#3}\rangle}} }

\def \ket#1{\mathinner{|{#1}\rangle}}
\def \bra#1{\mathinner{\langle{#1}|}}

\begin{document}

\title[Functional integral approach to heat exchange]{Functional Integral approach to time-dependent  heat exchange in open quantum systems:
general method and applications}

\author{{\bf M Carrega$^{1}$, P Solinas$^{1}$, A Braggio$^{1}$, M Sassetti$^{2,1} $, and U Weiss$^{3}$}}
\address{$^1$ SPIN-CNR, Via Dodecaneso 33, 16146 Genova, Italy}
\address{$^2$ Dipartimento di Fisica, Universit\`a di Genova, Via Dodecaneso 33, 16146 Genova, Italy}
\address{$^{3}$ II. Institut f\"ur Theoretische Physik, Universit\"at Stuttgart,
             D-70550 Stuttgart, Germany}

\begin{abstract}
We establish the  path integral approach for the time-dependent heat exchange of an externally driven quantum system coupled to a thermal reservoir. We derive the relevant influence functional and present an exact formal expression for the moment generating functional which carries all statistical properties of the heat exchange process for general linear dissipation. The general method is applied to the time-dependent average  heat transfer in  the dissipative two-state system. We show that the heat can be written as a convolution integral which involves the population and coherence correlation functions of the two-state system and additional correlations due to a polarization of the reservoir. The 
corresponding expression can be solved in the weak-damping limit both for white noise and for quantum mechanical coloured noise. The implications of pure quantum effects are discussed. Altogether a complete description of the dynamics of the average heat transfer ranging from the classical regime down to zero temperature is achieved.
\end{abstract}

\section{Introduction}
{
In recent years enormous advancements  have been accomplished in the fast growing field of quantum thermodynamics~\cite{esposito2009nonequilibrium, esposito2009erratum, campisi2011colloquium, campisi2011erratum}. 
This area of research is spurred both for fundamental and practical reasons.
On the fundamental side, one of the main question is how classical thermodynamics emerges from quantum physics~\cite{levy2012quantumPRE, Kosloff-Levy2014}.
On the  practical side,  there is an urgent need for a thorough understanding: how quantum systems exchange
energy and heat as a function of time? This is a prerequisite for building
quantum devices and/or quantum heat engines with potentially enormous technological impact~\cite{Kosloff-Levy2014, levy2012quantum}.

Regardless of the new  developments, quantum thermodynamics can still be considered as
a ``young'' field of research.
This is because there is an intrinsic difficulty in considering the concept of classical thermodynamics at quantum level since one has to face with many  problems.

A striking example is the attempt to generalize the definition of work and heat done and dissipated in a quantum system, which is still heavily debated~\cite{esposito2009nonequilibrium, campisi2009fluctuation}.
The most common approach is to measure the energy of the system twice and then iterate the protocol to reconstruct statistical information~\cite{kurchan2000quantum, tasaki2000jarzynski, talkner2007fluctuation, esposito2009nonequilibrium, campisi2011colloquium, dorner2013extracting, mazzola2013measuring, batalho2014}. Despite its simplicity and usefulness, this approach is limited by the fact that a standard projective measurement would lead to  wave function collapse and thus would seriously  perturb the system dynamics.
Alternative approaches have been proposed but the main issues are still open~\cite{solinas2013work, watanabe2014generalized}.

A complementary approach is to measure the energy, or heat, dissipated because of interactions between the system and an environment. In this case, {it is essential to study} the degrees of freedom of the environment itself. In this direction, a promising experimental proposal was put forward in Ref.~\cite{pekola2013calorimetric, gasparinetti2014fast}, where the setup consists in an engineered resistor acting as an environment with a sufficiently small heat capacity, allowing the measurement of the change in temperature when a quantum of energy is emitted or absorbed.

A second, and more technical, problem lies in the approximations generally made in describing the dynamics of a quantum system which interacts with an environment. In fact a  rigorous microscopic 
 description of energy exchange and dissipation is still lacking.

In the following, we assume the standard definition of heat~\cite{talkner2007fluctuation, esposito2009nonequilibrium, campisi2011colloquium,gasparinetti2014heat, silaev2014lindblad, wollfarth2014distribution} and focus on the second issue by  presenting a microscopic derivation  of  the process of  heat exchange of  a quantum system
with the surroundings.
We will approach the issue with  functional integral techniques.
In such a formalism, all quantum effects of the environment  on a microscopic  system, which are 
related to energy transfer, can be included in a specific influence functional of the coordinates of the system only.
Within the requirements of a linear environment all conceivable quantum noise sources may be accounted for
by appropriate  spectral properties of the influence kernels. 
The central problem of this study is to develop a general formalism for 
finding all statistical features of the energy transfer between the system and the environment.
In order to investigate the properties and advantages of this 
approach, and to map out potential of the method we will apply it to a specific quantum systems.

As in the context of dissipative quantum dynamics~\cite{weiss1999quantum, ingold2002path}, 
the influence functional does not depend on the particular quantum system under consideration. In addition, this
formalism allows the study of the full dynamics of the system and  supports   insights 
into the physical processes of emission, absorption, and dissipation of energy.
Moreover, one can take advantage of many numerical and non-perturbative  methods~\cite{orth2010universality, orth2013nonperturbative, henriet2014quantum, bulla2003numerical, egger1994low}, as well as
different  well-established analytical techniques~\cite{sassetti1990universality, grifoni1996exact, weiss1999quantum, egger1994low, keil2001real, grifoni1995cooperative, kopp2007universal}, all of them available on
the "market place" of dissipative quantum mechanics.

We will apply the general formalism to the simple but non-trivial dissipative two-state or 
spin-boson model~\cite{weiss1999quantum, grifoni1998driven, kehrein1996spin, grifoni1993nonlinear, grifoni1997dissipative, hartmann2000driven}. We will derive an exact formal expression for the time evolution of
the average heat and heat power in the presence of an external driving field. We will show that for a constant bias
the heat power can be expressed in the form of a  convolution which involves the  population and coherence correlation functions of the model system superimposed by polarization correlations of the heat bath accounting for heat exchange. We will present results in analytic form for weak Ohmic friction which cover both the white-noise and
the coloured  quantum noise regime.

In Sect.~\ref{sect:model} we will present  the microscopic model and introduce the  characteristic function
which includes the entire statistics of the energy transfer process. Specifically, we will define the characteristic 
function as the trace of  a  generalized density matrix of the system-plus-bath entity. 
This will  enable us perform the reduction of the generating functional to the coordinates of the system alone by
a suitable generalization of the Feynman-Vernon method. The reduction will be performed  for general coloured quantum noise in Sect.~\ref{sect:path}.
The explicit expression for the influence functional generalizes the Feynman-Vernon functional. While  the latter accounts for decoherence and relaxation of the reduced system, the generalized version captures also the quantum effects of the environment upon the time-dependent energy transfer process. To investigate potential of the method, we will consider in Sect.~\ref{sect:tls} the average heat in the dissipative two-state system and we will present an exact formal expression suitable  as a source for performing dynamical simulations in the presence of external driving. In addition, we will give  exact expressions in analytic form for a constant bias and weak Ohmic friction which cover the
entire temperature domain ranging from the classical white noise regime down to zero temperature where coloured
quantum noise prevails. We point out that these effects could be observed using a setup similar to the one proposed in Ref.~\cite{pekola2013calorimetric, gasparinetti2014fast}, provided that the detector has {sensitivity} to resolve in real time the dynamics of the system at {sufficiently} low temperatures. 
Finally, in Sect~\ref{sect:concl} we will present our conclusions.

\section{Model and general settings}
\label{sect:model}
We consider a { quantum mechanical particle of mass $M$,  position $q$ and momentum $p$ moving in a time-dependent potential  $V(q,t)$}. We assume that the particle is coupled to a thermal bath, modelled as a set of $N$ independent harmonic oscillators~\cite{caldeira1983path, weiss1999quantum, ingold2002path, leggett1987dynamics}. 
The Hamiltonians of the system S and reservoir R  are
\begin{eqnarray}
\label{eq:hs}
H_{\rm S} &=&  \frac{p^2}{2 M } + {   V(q,t)     }     \, ,\\
H_{\rm R} &=& \sum_{\alpha=1} ^N \Big [ \frac{p_\alpha^2}{2 m_\alpha } + \frac{1}{2} m_\alpha \omega^2_\alpha x_\alpha^2 \Big ]~,
\label{eq:hr}
\end{eqnarray}
where  $x_{\alpha}$ and $p_{\alpha}$ are position and momentum of the harmonic oscillator $\alpha$, respectively.   
The system-reservoir coupling is chosen to be bi-linear in the positions variables of system and reservoir. 
 It reads 
\begin{equation}
 H_{\rm I} = -q \sum_{\alpha=1} ^N c_\alpha x_\alpha +\frac{1}{2} \sum_{\alpha=1}^N \frac{c_\alpha^2}{m_\alpha \omega_\alpha^2} q^2~,
\label{eq:Hint}
\end{equation}
with $c_\alpha$ the interaction strength. 
The second term in $H_{\rm I}$ has been added to compensate renormalization of the  potential $V(q)$  caused by  the bi-linear coupling term in $H_{\rm I}$~\cite{weiss1999quantum, ingold2002path}.
All effects of the reservoir on the system induced by the coupling $H_{\rm I}$ 
are captured by the spectral density of the coupling 
\begin{equation}
\label{spectraldefinition}
 J(\omega) = \frac{\pi}{ 2} \sum_{\alpha=1} ^N \frac{c_\alpha^2}{m_\alpha \omega_\alpha} \delta (\omega -\omega_\alpha)~.
\end{equation}

Being interested in the continuum limit $N\to\infty$, we consider smooth
spectral densities of the coupling with power-law behaviour at low frequencies and spectral cut-off, e.g., exponential cutoff,  at high frequencies,
\be
\label{eq:Jspec}
J(\omega ) = M\gamma_s \omega \left(\frac{\omega}{\tilde\omega}\right)^{s-1}e^{- \omega/\omega_{\rm c}}~ .
\ee
Here, $\gamma_s$ is the coupling constant, $\tilde\omega$ a reference frequency and $\omega_{\rm c}$ the 
high-frequency cut-off.
The important case of Ohmic damping is represented by $s=1$ with $M\gamma_1\equiv \eta$ being the viscosity coefficient. The regime $s >1$ ($0< s <1$) describes super- (sub-) Ohmic dissipation~\cite{gorlich1989low, weiss1999quantum, grifoni1998driven, grifoni1997dissipative}. 
{In the sequel we consistently use units in which $k_{\rm B}^{}= \hbar=1$.}

Our goal is to determine  the amount and the statistics of the exchanged energy 
between the system and the reservoir  in a time interval $t$ (hereafter we put the initial time $t_0=0$).
This information may be obtained by performing a projective  measurement of the energy of the environment both at the beginning and at the end of the time interval.
Different kinds of the preparation of the initial state are feasible. 
In the following,  we consider the density matrix of the composite system at time zero in factorized form~\cite{weiss1999quantum},  
\begin{equation}
\rho(0) = \rho_{\rm S}(0) \otimes \rho_{\rm R}(0)\, .
\label{eq:fac}
\end{equation} 
We assume that $\rho_{\rm R}(0)$ is the canonical density of the unperturbed reservoir,
\begin{equation}\label{eq:rhofree}
 \rho_{\rm R}(0) = \frac{1}{Z_{\rm R}} e^{- H_{\rm R}/T } \, ,
 \end{equation}
where $Z_{\rm R}$ is its partition function.
{  Then}  $H_{\rm R}$ commutes with $\rho(0)$, and thus the initial measurement 
of the heat of the reservoir does not affect the dynamics.
With this assumption, we can formulate  the conditional probability $P(Q,t)$ for the output of the two measures and, therefore, for the dissipated heat $Q$~\cite{gasparinetti2014heat}.

The central quantity of interest, {which includes the entire statistics of the heat transfer process,
is the characteristic function $G_\nu(t)$  defined  as  Fourier transform of the probability distribution $P(Q,t)$,  }
\be
G_\nu(t) = \int_{-\infty}^{+\infty} {\rm d}Q\,P(Q,t) \,{e}^{i \nu  Q}  = \sum_{n=0}^\infty \frac{(i \nu)^n }{n!} 
\langle\, Q^n(t) \,\rangle \, .
\ee
Because of the second equality, the function $G_\nu(t)$ is referred to as the moment generating function (MGF).

The $n$-th derivative of the MGF at $\nu=0$ gives the $n$th moment,    
\be
\label{eq:moments}
\average{Q^n(t)}= \left.(-i)^n \frac{d^n }{d \nu^n}G_\nu(  t)\right|_{\nu=0} \, .
\ee
The first moment, $n=1$, yields the average heat transferred to the reservoir~\cite{gasparinetti2014heat},
\be
\label{eq:moment_1}
\average{Q (t) }=\left.-i \frac{d G_\nu (t)}{d \nu}\right|_{\nu=0}~.
\ee

A major advantage of using the MGF is that it can be written as the 
trace of  a generalized density operator. Thus,  known techniques can be applied  to solve the dynamics. 
Following Ref.~\cite{gasparinetti2014heat}, we {may}  write
\begin{equation}
 \CGF(t) = \Trace [e^{i \nu H_{\rm R}} U(t, 0) e^{-i \nu H_{\rm R}} \rho(0)U^\dagger (t, 0) ] \, ,
 \label{eq:1}
\end{equation}
where the trace is over all degrees of freedom and $U(t, 0)$ is the {unitary evolution operator of the composite system,
i.e., the corresponding generator is the total Hamiltonian $H(t)=H_{\rm S}(t) +H_{\rm R}+H_{\rm I}$.}

In Ref.~\cite{gasparinetti2014heat} the MGF has been studied  within the weak-coupling Born-Markov approximation
 by solving a master equation associated with a generalized density matrix of the total system.
In this work we will advance significantly  by investigating the MGF $G_\nu(t)$ from a quite different perspective. In essence, we shall derive a path integral formulation for the characteristic function $G_\nu (t)$  which may serve as a basis for the study of all statistical moments of the heat transfer process, as specified in Eq.~(\ref{eq:moments}).\\

We do not specify a priori the particular form of the initial state $\rho_{\rm S}(0)$. The only assumption we make for convenience is that the system-bath complex is initially in a product state and that the 
system-bath interaction is switched on at  time $t=0^+$. However, the approach envisaged here can easily be generalized to other scenarios of the initial preparation.

\section{Path integral representation of the characteristic generating function $\CGF (t)$}
\label{sect:path}
In this section we take the route of obtaining  a closed formal expression for the characteristic function $G_\nu(t)$.
It is advantageous to carry out the reduction of the dynamics of the system-plus-reservoir entity to the dynamics of the system alone within the functional integral description~\cite{weiss1999quantum, ingold2002path}, a technique introduced by Feynman and Vernon already in 1963~\cite{feynman1963theory}.
The present problem, however,  is substantially different because of the additional operator terms ${e}^{i \nu H_{\rm R}}$ and
${e}^{-i \nu H_{\rm R}}$ in the original expression (\ref{eq:1}). 

The coordinate representation of Eq.~(\ref{eq:1}) reads
\begin{equation}
\label{eq:pos_g}
 \CGF(t) = \int d{\boldsymbol x}'_f \int dq_f  \bra{{\boldsymbol x}'_f, q_f} e^{i \nu H_{\rm R}} U(t, 0) {\rho}^{(\nu)}(0) 
U^\dagger (t, 0)  
\ket{ {\boldsymbol x}_f', q_f}  \, ,
\end{equation} 
where the $N$-component vector ${\boldsymbol x}'_f$ stands for  the bath coordinates $(x'_{1f}, ....,x'_{Nf})$, 
and $q_f$ represents a given position of the particle.
For convenience,  the factor ${e}^{-i \nu H_{\rm R}}$  in Eq.~(\ref{eq:1}) is included in the initial density matrix of the reservoir
\begin{equation}\label{eq:inicond}
{\rho}^{(\nu)}(0) = \rho_{\rm S}(0) \otimes {\rho}^{(\nu)}_{\rm R} (0)   
\end{equation}
with the shifted canonical density operator of the reservoir
\be
{\rho}^{(\nu)}_{\rm R} (0) = \frac{1}{Z_{\rm R}}{\rm  e}^{- H_{\rm R}/T  -i\, \nu H_{\rm R}}\, .
\label{eq:canonshifted}
\ee
Using completeness  of the position eigenstates we can rewrite $\CGF(t)$ as 
\beq
\label{eq:pos_g1}
 \CGF(t) &=& \int d{\boldsymbol x}_f d{\boldsymbol x}'_f d{\boldsymbol x}_i d{\boldsymbol x}'_i  \int dq_f dq_i dq'_i
 \nonumber \\
 &\times & 
\langle {{\boldsymbol x}'_f}|{ {e}^{i \nu H_R}}|{{\boldsymbol x}_f}\rangle
 K({\boldsymbol x}_f, q_f, t; {\boldsymbol x}_i, q_i, 0) 
\langle{{\boldsymbol x}_i q_i}|{{\rho^{(\nu)}}(0)}|{{\boldsymbol x}'_i q'_i}\rangle 
K^*({\boldsymbol x}'_f, q_f, t; {\boldsymbol x}'_i, q'_i, 0)~.
\eeq
Here, 
\begin{equation}
 K({\boldsymbol x}_f, q_f, t; {\boldsymbol x}_i, q_i, 0) \equiv \langle{{\boldsymbol x}_f, q_f}|{ U(t,0)}|{{\boldsymbol x}_i, q_i}\rangle
\end{equation}
is  the probability amplitude for the transition from the  initial state $i$ to the final state$f$ according to evolution with $U(t,0)$. 
The path integral representation of the propagator is~\cite{weiss1999quantum, feynman1963theory, ingold2002path}
\begin{equation}
  K({\boldsymbol x}_f, q_f, t;{\boldsymbol x}_i, q_i, 0) =\int\limits_{q(0)=q_i}^{q(t)=q_f} \!\!\!\!\! \D q \,{e}^{i S_{\rm S}[q]} \int\limits_{{\boldsymbol x}(0)= {\boldsymbol x}_i}^{{\boldsymbol x}(t)={\boldsymbol x}_f}\!\!\! \!\!
\D {\boldsymbol x} \,{e}^{ i ( S_{\rm R}[{\boldsymbol x}]  + S_{\rm I}[q,{\boldsymbol  x}]  )  }  \, .
\end{equation}
Here, $S_{\rm S}[q]$ is the action of the system, $S_{\rm R}[{\boldsymbol x}]$ the action of the reservoir, and 
$S_{\rm I}[q, {\boldsymbol x}]$ is the action associated with the interaction $H_{\rm I}[q,{\boldsymbol x}]$. With the
factorizing form (\ref{eq:inicond}) of ${\rho}^{(\nu)}(0)$ one then obtains 
\begin{equation}
\CGF(t) = \int dq_f dq_i dq'_i  \matrixel{q_i}{\rho_{\rm S}(0)}{q'_i}  J^{(\nu)} (q_f,t; q_i,q'_i,0) \, .
\label{eq:Grepres}
\end{equation}
The propagating function $J^{(\nu)}$  describes the time evolution of the RDM
resulting from the generalized density operator given in Eq.~(\ref{eq:1}) upon integrating out the reservoir. It reads
\begin{equation}\label{eq:propfunc}
 J^{(\nu)} (q_f,t; q_i,q'_i,0) =\int\limits_{q(0)=q_i}^{q(t)=q_f}\!\!\D q\, e^{i S_{\rm S}[q]}
\int\limits_{q'(0) = q'_i}^{q'(t)=q_f} \!\!\!\!\D q' \,e^{-i S_{\rm S}[q']} \mathcal{F}^{(\nu)} [q,q'] \, .
\end{equation}
For $ \mathcal{F}^{(\nu)} [q,q']=1$, $ J^{(\nu)} (q_f,t; q_i,q'_i,0) $ describes propagation of the RDM in the absence of the system-reservoir interaction $H_{\rm I}$. This illustrates that the functional $ \mathcal{F}^{(\nu)} [q,q']$ is the influence functional
for the characteristic function $G_\nu(t)$. This functional carries all effects of the system-reservoir coupling
{both on the dynamics of the reduced system and on the heat exchange process with the environment.} The still formal expression reads
\begin{equation}
 \mathcal{F}^{(\nu)} [q,q']= \int d{\boldsymbol x}_f d{\boldsymbol x}'_f d{\boldsymbol x}_i d{\boldsymbol x}'_i 
\matrixel{{\boldsymbol x}_i}{{\rho}^{(\nu)}_{\rm R}(0)}{{\boldsymbol x}'_i}
 \matrixel{{\boldsymbol x}'_f}{e^{i \nu H_{\rm R}}}{{\boldsymbol x}_f} F[q,{\boldsymbol x}_f, {\boldsymbol x}_i] 
F^*[q', {\boldsymbol x}'_f, {\boldsymbol x}'_i]  \, .
\label{eq:influence}
 \end{equation}
Here the amplitude
\begin{equation}
 F[q,{\boldsymbol x}_f, {\boldsymbol x}_i] =\int\limits_{{\boldsymbol x}(0)=
{\boldsymbol x}_i}^{{\boldsymbol x}(t)={\boldsymbol x}_f} \!\! \D{\boldsymbol  x} \,
{e}^{i\, ( S_{\rm R}[{\boldsymbol x}]+S_{\rm I}[q,{\boldsymbol x}]  )} 
 \ee
is the propagator of the bath in presence of the system-bath coupling {$H_{\rm I}$ given in Eq.~(\ref{eq:Hint}).}
In the limit $\nu\to 0$, the matrix element  $\matrixel{{\boldsymbol x}'_f}{e^{i \nu H_{\rm R}}}{{\boldsymbol x}_f}$ would
reduce to $\prod_\alpha \delta( x'_{\alpha f}-x_{\alpha f} )$ and hence $ \mathcal{F}^{(\nu)} [q,q']$ to the standard Feynman-Vernon
influence functional $\mathcal{F}^{(\nu=0)}[q,q'] =\mathcal{F}_{\rm FV} [q,q']$. 
 
The expression in Eq.~(\ref{eq:Grepres}) is formally similar to the one obtained for the RDM $\rho_{\rm S}(t)$ reported in~\cite{weiss1999quantum} in the context of quantum dissipative system. However, the substantial  difference is
the extra factor $\bra{{\boldsymbol x}'_f } e^{i \nu H_{\rm R}} \ket{{\boldsymbol x}_f}$ in  $\mathcal{F}^{(\nu)}[q,q']$ and $\rho_{{\rm R}}^{(\nu)}$ which entail independent integrations of  the end/initial  point ${\boldsymbol x}'_{f/i}$ and  ${\boldsymbol x}_{f/i}$.

It is convenient to express the influence functional (\ref{eq:influence})     in terms of the linear combinations
\be  \label{etaxipaths}
\eta(\tau) = [q(\tau)+ q'(\tau)]/ q_0 \, ,\qquad\mbox{and}\qquad \xi(\tau) = [q(\tau)- q'(\tau)]/q_0 \, .
\ee
Here, $q_0$ is a characteristic length of the system S, which is introduced to {render 
the paths $\eta(\tau)$ and $\xi(\tau)$ dimensionless.}
The path $\eta(\tau)$, usually referred to as quasi-classical, measures propagation of the system along the diagonal of the density matrix, whereas the path $\xi(\tau)$ measures off-diagonal excursions, i.e., quantum fluctuations.

In order to evaluate Eq.~(\ref{eq:influence}) we follow the standard approach~\cite{weiss1999quantum, grifoni1998driven} of gaussian path integral integrations. 
The details of the calculation are presented in \ref{app:gaussianintegration}.
With the resulting expressions  (\ref{eq:mfl_alpha}) and  (\ref{eq:Finflapp}) -- (\ref{eq:kernels_l}) , the  influence functional is found to read
\be    \label{eq:mfl}
 \mathcal{F}^{(\nu)}[\eta, \xi] =   {e}^{ -(\,\Phi^{(0)}_{\rm loc}[\eta, \xi] 
 +\Phi^{(0)}_{\rm cor}[\eta, \xi] \, ) } \;
 {\rm  e}^{ i\, \Delta \Phi^{(\nu)}[\eta, \xi] }~ .
 \ee
The influence actions of the first exponential factor are
\begin{equation}  \label{eq:PhiFV}
\begin{array}{rcl}
 \Phi^{(0)}_{\rm loc}[\eta, \xi] &=& {\displaystyle 
\;  i\, \frac{\mu q_0^2}{2} \int_0^t \!dt' \,\xi(t') \eta (t')  
\quad\;\; \mbox{with}\quad\;\; \mu= \frac{2}{\pi} \int_0^\infty \!\!d\omega\,\frac{J(\omega)}{\omega}  \, ,  }  \\[4mm]
 \Phi^{(0)}_{\rm cor}[\eta, \xi] &=& {\displaystyle
\int_0^t dt' \!\int_0^{t'}\!\! dt'' \,  \xi(t') [L'(t'-t'') \xi(t'') + i\,  L'' (t'-t'') \eta (t'') ]  \, . } 
\end{array}
\end{equation}
The local action $ \Phi^{(0)}_{\rm loc}[\eta, \xi]$ originates from the potential counterterm in the interaction (\ref{eq:Hint}).  The nonlocal term $\Phi^{(0)}_{\rm cor}[\eta, \xi] $ is the standard Feynman-Vernon influence action functional which governs decoherence and relaxation of the reduced system~\cite{weiss1999quantum}.
The real part of $ \Phi^{(0)}_{\rm cor}[\eta, \xi]$  describes time-correlations of the quantum fluctuations, and 
the imaginary part describes time-correlations between quasiclassical propagation and
subsequent off-diagonal excursions. The kernel $L(\tau)$ is the equilibrium time-correlation function of the
collective bath mode ${\cal E}(t) = q_0\sum_\alpha c_\alpha x_\alpha(t)$,
\be  \label{eq:polcorr}
L(\tau)  =  L'(\tau) + i L''(\tau) = \frac{q_0^2}{\pi} \int_0^{\infty}  \!\!\! d\omega\,  J(\omega) \Big[\,\coth 
\Big(\frac{ \omega}{2 T}\Big)  \cos \omega\tau - i \sin \omega\tau\, \Big] \,  .
\ee
The imaginary part $L''(\tau)$ is proportional to the time derivative of the friction kernel 
$M\gamma(\tau)$ in the associated classical equation of motion ($M$ is the inertial mass of the system),
\be\label{eq:gammarel}
\Theta(\tau) \, L'' (\tau) = \frac{M q_0^2}{2} \, \dot{\gamma}(\tau)  \, .
\ee

The second exponential factor in Eq.~(\ref{eq:mfl}) controls the statistics of the heat transport process between the system and the environment,
\begin{eqnarray}
\Delta \Phi^{(\nu)}[\eta, \xi]&=& \int_0^t \!dt'\!\!\int_0^{t' }\!\!dt''\!
\left\{ \big[ \eta(t') \eta(t'') - \xi(t') \xi(t'')\big] L^{(\nu)}_1 (t'-t'')\right.     \nonumber\\
&& \qquad  \qquad \quad \left. +\;\big[ \eta(t')\xi(t'') - \xi(t') \eta(t'') \big] L^{(\nu)}_2 (t'-t'')\right\}   \, .
\label{eq:deltaphifin} 
\end{eqnarray}
The influence action $\Delta \Phi^{(\nu)}[\eta, \xi]$ carries in addition to the  time-correlations of the aforementioned types also time-correlations between quasiclassical propagation at different time
and between quantum fluctuations and subsequent quasiclassical propagation.
The spectral representations of the kernels are
\begin{equation}   \label{eq:Lnu12}  
\begin{array}{rcl}
L^{(\nu)}_1 (\tau) &=&\; {\displaystyle \frac{q_0^2}{\pi} \int_0^{\infty}  \!\!\! d\omega\, 
\frac{\sin(\nu\,\omega/2) \sinh[(\omega/T+ i\,\nu\,\omega)/2]}{\sinh[\omega/(2T)]}\,J(\omega)\cos(\omega \tau)  }
\, ,    \\[5mm]
L^{(\nu)}_2 (\tau) &=&{\displaystyle -\,i\, \frac{q_0^2}{\pi} \int_0^{\infty}  \!\!\! d\omega\,  
\frac{\sin(\nu\,\omega/2) \cosh[(\omega/T+ i\,\nu\,\omega)/2]}{\sinh[\omega/(2T)]}\,J(\omega)\sin(\omega \tau)   }
\, .
\end{array}
\end{equation}

 As $\nu\to 0$, the functional ${e}^{ i\Delta \Phi^{(\nu)}[\eta, \xi] } $ approaches unity.
With the influence functional (\ref{eq:mfl})  and the shortform
$S_{\rm S}[\eta,\xi] = S_{\rm S}[q] - S_{\rm S}[q']$ the moment generating function now reads
\be  \label{eq:cgffinal}
\fl\qquad  G_\nu(t) = \int d\eta_i \,\bra{  \eta_i} \rho_{\rm S}(0) \ket{\eta_i} \int d\eta_f   \!\!\!
\int\limits_{\eta(0) = \eta_i}^{\eta(t)=\eta_f}  \!\!\! \D \eta \int\limits_{\xi(0)=0}^{\xi(t)=0}
\!\!\!    \D \xi\;
{e}^{i\,S_{\rm S}[\eta,\xi]}\;
  {e}^{ -( \Phi^{(0)}_{\rm loc}[\eta, \xi] + \Phi^{(0)}_{\rm cor}[\eta, \xi] )}  \;
 {\rm  e}^{ i\, \Delta \Phi^{(\nu)}[\eta, \xi] }  \, .
\ee
Here we have assumed that the system is initially in a diagonal state of its density matrix.

The time-dependent average heat transfer between the system and the reservoir is now obtained according to the
relation (\ref{eq:moment_1}) as
\be  \label{eq:avheatfinal}
\fl \qquad \average{Q(t)} = \int d\eta_i \,\bra{  \eta_i} \rho_{\rm S}(0) \ket{\eta_i} \int d\eta_f   \!\!\!
\int\limits_{\eta(0) = \eta_i}^{\eta(t)=\eta_f}  \!\!\! \D \eta \int\limits_{\xi(0)=0}^{\xi(t)=0}
\!\!\!    \D \xi\;  \Phi^{(1)}[\eta,\xi]\; 
{e}^{i\,S_{\rm S}[\eta,\xi]} \;
  {e}^{ -( \Phi^{(0)}_{\rm loc}[\eta, \xi]  +  \Phi^{(0)}_{\rm cor}[\eta, \xi] ) } \, ,
\ee
where $\Phi^{(1)}[\eta,\xi] = \frac{d}{d\nu} \Delta\Phi^{(\nu)}[\eta,\xi]\Big|_{\nu = 0}$. The resulting expression for
$\Phi^{(1)}[\eta,\xi]$ is
\begin{equation} \label{eq:Phi1}
\begin{array}{rcl}
\Phi^{(1)}[\eta, \xi] &=&{\displaystyle \frac{1}{2}  \int_0^t \!dt'\!\!\int_0^{t' }\!\!dt''\!
\left\{ \big[ \xi(t') \xi(t'') -  \eta(t') \eta(t'') \big] \dot{L}''(t'-t'')\right.  } \nonumber\\[4mm]
&& \qquad  \qquad \quad \left. +\,i\, \big[ \eta(t')\xi(t'')   -  \xi(t') \eta(t'')  \big] \dot{L}'(t'-t'')\right\}  \, ,
\end{array}
\end{equation}
in which $\dot{L}'(\tau)$ and $\dot{L}''(\tau)$ are  real and imaginary part of the time-derivative of the
correlation function  $L(\tau)$ given in Eq.~(\ref{eq:polcorr}).
The explicit expressions (\ref{eq:cgffinal}) and (\ref{eq:avheatfinal})  for the MGF $G_\nu(t)$ and the
average heat transfer $\average{Q(t)}$ with the expressions (\ref{eq:deltaphifin}) and (\ref{eq:Phi1}) for
$\Delta\Phi^{(\nu)}[\eta,\xi]$ and $\Phi^{(1)}[\eta,\xi]$, respectively,
are the main results of this section. They  hold for general linear dissipation with
any type of memory-friction.

 We conclude this section with a short additional consideration. Firstly,  we see that the term 
 in (\ref{eq:Phi1})  which describes time correlations between quasiclassical paths 
 yields a contribution to $\average{Q(t)}$ even if
the system is frozen in its initial position.
The constant path $q(\tau)= q'(\tau) = \frac{1}{2} \eta_i q_0$ yields the contribution
\be\label{eq:avdeltaQ0}
\average{\delta Q(t)} = \int d\eta_i \, p(\eta_i) \average{\delta Q_i(t)}
\qquad \mbox{with}\qquad
 \average{\delta Q_i(t)} = \Phi^{(1)}[\eta_i,0]  \, .
\ee
Here $p(\eta_i)= \bra{  \eta_i} \rho_{\rm S}(0) \ket{\eta_i} $ is the occupation probability 
of the initial state $\eta_i$,
upon using the relation (\ref{eq:gammarel}), we can relate the heat portion $\average{\delta Q_i(t)} $ to the 
friction kernel $M \gamma(t)$,

\be
 \average{\delta Q_i(t)} = \frac{ \eta_i^2 q_0^2}{4}  M \,  \big[\, \gamma(0^{+}) - \gamma(t)\, \big] \, .
\ee
Remarkably, the heat portion (\ref{eq:avdeltaQ0})  generally occurs when the system dynamics is slow
on the scale $1/\omega_{\rm c}$.

 Evidently, for the thermal initial state  (\ref{eq:rhofree}) of the uncoupled reservoir, the  heat contribution 
 $\average{\delta Q_i(t)}$ depends on the particular initial state $\eta_i$ of the system. In fact, when  the coupling
$H_{\rm I}[\eta_i q_0/2,{\boldsymbol x}]$ is switched on at time zero, the thermal state of the uncoupled 
reservoir relaxes to a shifted canonical state.
The initial slip (\ref{eq:avdeltaQ0}) on the time scale $1/\omega_{\rm c}$ represents the respective heat transfer.
If we had calculated the functional $\Phi^{(1)}[\eta,\xi]$ for a canonical initial state with
shifted oscillator positions, $x_\alpha -   \bar{\eta} q_0 c_\alpha/(2 m_\alpha \omega_\alpha^2)$, and had subtracted 
the irrelevant constant polarization energy $ (\bar{\eta} q_0)^2 \,\mu/8$, the heat contribution would be
\be \label{eq:avdeltaQ1}
\average{\delta Q_{i}(t)} = \frac{q_0^2}{4} \, \eta_i(\eta_i- \bar{\eta})  M \, 
\big[\, \gamma(0^{+} )-\gamma(t) \, \big]  \, .
\ee
With the choice $\bar{\eta} =\eta_i$ the initial slip term (\ref{eq:avdeltaQ1})  would be absent. The relevant shifted thermal initial state can be arranged, e.g., by trapping the system at times $t<0$ in the particular state
$\eta(t)=\eta_i$ and releasing the constraint at time zero. Keeping this in mind,  we disregard the 
initial slip contribution in the sequel.

\section{Application to the spin-boson model: the average heat transfer}
\label{sect:tls}

Consider a particle moving in a double well potential and in contact with its environment and assume that the thermal energy is small compared to the energy splitting in the individual wells. Then only
the two lowest states become relevant to the dynamics.
This scenario  is conveniently described by the spin-boson model. It is of great interest in different fields 
ranging from nuclear magnetic systems to 
quantum optics~\cite{grifoni1998driven, orth2010universality, orth2013nonperturbative, kehrein1996spin}.

We now apply the method developed in the previous section to the driven 
spin-boson model~\cite{weiss1999quantum, grifoni1998driven, kehrein1996spin}. Here our interest is focused 
on the transferred average heat and on the heat power.
The Hamiltonian of the two-state system (TSS) in pseudo-spin representation is
\begin{equation}
\label{eq:sb}
H_{\rm S}=-\frac{\Delta}{2} \sigma_x - \frac{\epsilon(t)}{2} \sigma_z~.
\end{equation}
The state basis is formed by the two localized states $\ket{R}$ and $\ket{L}$ of the double well, which are eigenstates of $\sigma_z$ with eigenvalues $+1$ and $-1$, respectively. The position  operator is  $q= q_0  \sigma_z/2$ and has
eigenvalues $\pm q_0/2$ being associated with the minima of the double-well potential $V(q)$. The tunneling operator 
$\sigma_x =\ket{R}\bra{L} -\ket{L}\bra{R}$ transfers the particle between the two wells with tunneling amplitude $\Delta$, while 
\begin{equation} \label{eq:bias}
\epsilon(t) = \epsilon_0 + \epsilon_1(t)
\end{equation}
describes an externally  applied bias with a constant part $\epsilon_0$, and a driving term $\epsilon_1(t)$. 

 The MGF expression in Eq.~(\ref{eq:cgffinal})  takes  for the TSS  in the $\{ \xi,\eta\}$-representation  the form
\begin{equation}  \label{eq:cgftss}
\fl \qquad
G_\nu(t) = \sum_{\eta_f=\pm 1}\sum_{\eta_i=\pm 1} \matrixel{\eta_i}{\rho_{\rm S}(0)}{\eta_i} 
\int_{\eta(0)=\eta_i}^{\eta(t) = \eta_f } \mathcal{D} \eta   \int_{\xi(0)=0}^{\xi(t)=0} \mathcal{D} \xi \,   
\,{e}^{i\, S_{\rm S}[\eta,\xi]} \, {e}^{- \Phi^{(0)}_{\rm cor}[\eta,\xi ]} 
\,{e}^{i\,\Delta\Phi^{(\nu)}[\eta,\xi ] }  \, .
\end{equation}
The corresponding expression for the transfer of heat reads
\begin{equation}  \label{eq:heattss}
\fl  \qquad
\average{Q(t)} = \sum_{\eta_f=\pm 1}\sum_{\eta_i=\pm 1} \matrixel{\eta_i}{\rho_{\rm S}(0)}{\eta_i} 
\int_{\eta(0)=\eta_i}^{\eta(t) = \eta_f } \mathcal{D} \eta   \int_{\xi(0)=0}^{\xi(t)=0} \mathcal{D} \xi \,   
\Phi^{(1)}[\eta,\xi]\,
\,{e}^{i\, S_{\rm S}[\eta,\xi]} \, {e}^{- \Phi^{(0)}_{\rm cor}[\eta,\xi ]}     \, .
\end{equation}
Here we have utilized that the local action $\Phi^{(0)}_{\rm loc}[\eta,\xi]$ 
defined in Eq.~(\ref{eq:PhiFV})  is zero for the TSS.

\subsection{Exact formal solution for the average heat }

The double path sum in Eq.~(\ref{eq:cgftss})  can be visualized as a path sum for a single path that successively visits the four states of the RDM, i.e., the two diagonal or sojourn states labelled by $\eta=\pm 1$,
and the two off-diagonal or blip states labelled by $\xi=\pm 1$. Starting out from a diagonal state of the RDM, 
 the path dwells in sojourn $j$ during time interval $s_j = t_{2j+1} - t_{2j} $, and in a blip during time interval 
$ \tau_j = t_{2j} - t_{2j-1}$, where $t_j$ is the flip time. The piecewise constant  paths
 $\eta(\tau)$- and $\xi(\tau)$  with $2n$ flips are given by
\begin{equation} \label{eq:etaxi}
\begin{array}{rcl}
\eta^{(n)}(\tau) &=&{\displaystyle 
\sum_{j=0}^n \eta_j \big[\,\Theta(\tau- t_{2j}) -  \Theta(\tau- t_{2j+1})\,\big]  \, , } \\[5mm]
\xi^{(n)}(\tau) &=& {\displaystyle
\sum_{j=1}^n \xi_j \big[\, \Theta(\tau- t_{2j-1}) -\Theta(\tau- t_{2j}) \,\big]  \, ,   }
\end{array}
\end{equation}      
where $t_0=0$ and $t_{2n+1} = t$. Here the initial state of the RDM is $\eta_i =\eta_0$ and the final state is
$\eta_f=\eta_n$.
The RDM at initial time $t=0$ is given by the initial populations 
$p_{ L}^{}= \frac{1}{2} (1+p_0)$ and $p_{ R}^{}= \frac{1}{2} (1-p_0)$ of the left and right well. We have
$-1 \le p_0 \le 1$ and
\be  \label{eq:ini}
\matrixel{\eta_0}{\rho_{\rm S}(0)}{\eta_0} = \frac{1}{2} (1-p_0)\,\delta_{\eta_0,1} + \frac{1}{2} (1+p_0)\, 
\delta_{\eta_0,-1} \, .
\ee

The path sum in the expressions (\ref{eq:cgftss})   results in a series in the
number of flip pairs, and in each term the  summation of all possible
arrangements in the time-ordered alternating series of sojourns and blips,
\begin{equation} \label{eq:sumetaxi}
\int \!\mathcal{D} \eta \int\! \mathcal{D} \xi \cdots \to\sum_{n=0}^{\infty} \int_0^t \!\! \mathcal{D}_{2 n} \{ t_j\}   
 \!\sum_{\{\xi_j=\pm 1\}} \sum_{\{\eta_j=\pm 1\}}\cdots\,  .
\end{equation}  
Here the integration symbol $\int\mathcal{D}_{2 n} \{ t_j\}  $ is a short form of the $2n$ time-ordered integrations,
\begin{equation}
 \int_0^t \! \mathcal{D}_{2 n} \{ t_j\} = \int_0^t   \!\! dt_{2n } \!\int_0^{t_{2n}}
\!\!\! dt_{2n-1} ....\int_0^{t_{2}}     \!\!\! dt_{1}  \, .
\end{equation}

Consider first the weight factors of the paths resulting from the internal dynamics of the 
TSS.
The weight to switch per unit time from the diagonal state $\eta_j $ to the directly following off-diagonal state 
$\xi_{j+1}$ is $-i\, \eta_j\,\xi_{j+1}\, \Delta/2$, and the weight to switch per unit time 
from the offdiagonal state $\xi_{j+1} $ to the directly following diagonal state $\eta_{j+1}$  is 
$-i\, \xi_{j+1}\,\eta_{j+1}\, \Delta/2$. Thus, the  aggregated  weight of $2n$ flips depends only on the initial and
final sojourn state,
\be   \label{eq:flipweight}
(-1)^n (\Delta^2/4)^n\, \eta_0\,  \eta_n  \, .
\ee
 The weight to stay in a sojourn is unity, whereas the weight to stay
in the blip $\xi_j$ during time interval $t_{2j}-t_{2j-1}$ is $\exp[\,i \xi_j \int_{t_{2j-1}}^{t_{2j}} dt'\epsilon(t')\, ]$.
Accordingly,  the contributions of $n$ blips  are accumulated  in the bias weight factor
\begin{equation}
B_n  = e^{i\, \varphi_n}, \qquad\varphi_n = \sum_{j=1}^n \xi_j\,
\Big[\epsilon_0  \tau_{j } + \int_{t_{2j-1}}^{t_{2j}} 
\!\!\! d\tau\, \epsilon_1(\tau) \Big] ~.
\end{equation}

For the piecewise constant path  (\ref{eq:etaxi}) in the influence action $\Phi_{\rm cor}^{(0)}[\eta,\xi]$  the time correlations 
among the $\{\eta_i\}$- and $\{\xi_j\}$- labels or charges
are expressed in terms of the second integral of the function $L(\tau)$,
\be   \label{eq:Qcorr}
W(\tau) =  W'(\tau) + i\,W''(\tau) = \frac{q_0^2}{\pi} 
\int_0^{\infty}  \!\!\! d\omega\,\frac{ J(\omega)}{\omega^2} \Big[ 
\coth \Big(\frac{\omega}{2 T}\Big) [\,1-\cos(\omega \tau)\,] + i\, \sin(\omega \tau) \,\Big]    \, .
\ee
The resulting correlation factor for $2n$ flips is~\cite{grifoni1998driven, grifoni1999dissipation, grifoni1997coherences}
\begin{eqnarray}
&&\Big(e^{-\Phi^{(0)}_{\rm cor}[\eta,\xi]}\Big)_n = G_nH_n   \, ,   \nonumber \\
&&G_n = \exp{\Big\{ - \sum_{j=1}^n  W'_{2j, 2j-1}  \Big\} }
\; \exp{\Big\{- \sum_{j=2}^n \sum_{k=1}^{j-1} \xi_j \Lambda_{j,k} \xi_k \Big\}}  \, , \\   \nonumber
&&H_n = \exp{ \Big\{ i  \sum_{j=1}^n \sum_{k=0}^{j-1} \xi_j X_{j,k} \eta_k \Big\}  }
= \exp\Big\{ i \, \sum_{k=0}^{n-1}  \phi_{nk} \eta_k \Big\}     \, ,  
\end{eqnarray}
where
\be  \label{eq:phikn}
\phi_{n,k} = \sum_{j=k+1}^n \xi_j X_{j,k}  \; .
\ee
The first exponential factor of $G_n$ represents the intrablip correlations, and the second exponential factor
the interblip correlations of the n blips.  With the short form  $W_{j,k} = W(t_j-t_k)$
the interactions between all sojourns and all subsequent blips are in the phase factor $H_n$.
In the second form, $H_n$ is written in terms of the entire blip correlations $\phi_{n,k}$ with sojourn $k$.
 The interactions between blip $k$ and subsequent blip $j$, and the interactions between
sojourn $k$ and subsequent blip $j$ are written as
\begin{equation}\label{eq:xjk}
\begin{array}{rcl}
\Lambda_{j,k} \!\!\!&=&\!\! \! W'_{2j,2k-1} + W'_{2j-1,2k}  - W'_{2j,2k} - W'_{2j-1,2k-1}   \,, 
\\[3mm]
X_{j,k}\!\!\! &=& \!\!\!  W''_{2j,2k+1} + W''_{2j-1,2k}  - W''_{2j,2k}  -  W''_{2j-1,2k+1}            \, .
\end{array}
\end{equation}

With these individual terms the average heat takes the form
\be   \label{eq:DeltaQ}
\begin{array}{rcl}
 {\average {Q(t)}} &=& {\displaystyle   \sum_{\eta_0=\pm 1} \matrixel{\eta_0}{\rho_{\rm S}(0)}{\eta_0}  \,
\eta_0 \,\sum_{n=1}^\infty \Big( \frac{-\Delta^2}{4}\Big)^{n} \int_0^t \mathcal{D}_{2 n} \{ t_j\}    }    \\[4mm]
&&\qquad   {\displaystyle  \times  \sum_{ \{\xi_j=\pm 1 \}}  B_n G_n \sum_{\{\eta_j =\pm 1\}''}  H_n
\sum_{\eta_n=\pm 1} \eta_n \, \Phi^{(1)}[\eta^{(n)},\xi^{(n)}]    \, .  }
 \end{array}
\ee
Here the double prime in  $\{\eta_j=\pm 1\}''$ indicates summation  over the internal sojourn states
$\eta_1,\cdots,\eta_{n-1}$. 

For the piecewise constant paths (\ref{eq:etaxi}) the functional $(\ref{eq:Phi1})$ describes time correlations between sojourns and blips with any time-ordering. By analogy with the relation of the kernels in the actions 
(\ref{eq:Phi1}) and (\ref{eq:PhiFV}),
the time correlations  between the $\{\eta_i\}$- and $\{\xi_j\}$-charges  in the  action
$\Phi^{(1)}[\eta^{(n)},\xi^{(n)}]$ are specified by the derivative of the function  $W(\tau)$ defined in Eq.~(\ref{eq:Qcorr}). 
The imaginary part $\dot{W}''(\tau)$ is connected with the damping kernel according to 
\be
\dot{W}''(\tau) =\frac{ M q_0^2}{2} \gamma(\tau) \,  .
\ee

Interestingly enough,  the summation over the final sojourn states $\eta_n=\pm 1$ in the expression (\ref{eq:DeltaQ})
cancels, by reason of the extra  factor $\eta_n$, all correlations of blips and sojourns in $\Phi^{(1)}$ except the 
correlations with the two final sojourns states. Summation of the leftover correlations yields
\begin{equation}
\sum_{ \eta_n=\pm 1 } \eta_n\Phi^{(1)}[\eta^{(n)},\xi^{(n)}] =  \sum_{k=0}^{n-1} \mathcal{U}_{n,k} (t) \,\eta_k
+i  \sum_{j=1}^{n} \mathcal{V}_{n,j} (t)\,  \xi_j \, .
\label{eq:Kappa} 
\end{equation}
Here  $\mathcal{U}_{n,k} (t)$ represents the correlations of sojourn $k$ with sojourn $n$ and
$\mathcal{V}_{n,j} (t) $ the correlations   of blip $j$ with sojourn $n$. Explicit dependence on the end time $t$
is indicated for later purpose. We obtain analogous to the forms  (\ref{eq:xjk})  the expressions
\be
\begin{array}{rcl}
{\mathcal U}_{n,k} (t) &=& \dot{W}''(t-t_{2k+1}) + \dot{W}''(t_{2n}-t_{2k})   - \dot{W}''(t-t_{2k}) - 
\dot{W}''(t_{2n}-t_{2k+1})\, , \\[4mm]
{\mathcal V}_{n,k}(t) &=&  \dot{W}'(t-t_{2k-1})+\dot{W}'(t_{2n} -t_{2k})-\dot{W}'(t-t_{2k})- \dot{W}'(t_{2n}-t_{2k-1}) \, .
\end{array}
\ee
Next, we add up  the contributions of the $2n-1$ intermediate sojourns. Further, we utilize that 
in Eq.~(\ref{eq:DeltaQ})  the contributions of the blip states which are odd under 
the substitutions $\{\xi_j\} \to \{-\xi_j\}$ cancel each other out.
The resulting expression for $ {\average {Q(t)}} $  is conveniently divided into the ${\cal U}$- and 
${\cal V}$-contributions, $\average{Q(t)}_1$ and $\average{Q(t)}_2$,   
and these into those which are symmetric (s)  and antisymmetric (a) under bias inversion ($\epsilon \to -\epsilon$).
We readily find
\be  \label{eq:Qsplit1}
\average{Q(t)} = \average{Q(t)} _1 + \average{Q(t)}_2  \, ,
\ee
where 
\be  \label{eq:Qsplit2}
\fl\qquad  {\average {Q(t)}}_r = \frac{1}{2}\, 
\sum_{n=1}^\infty  \Big( \frac{-\Delta^2}{2}\Big)^n  
\int_0^t \mathcal{D}_{2n} \{t_j\} \sum_{\{ \xi_j=\pm 1\}}  \!   
\bigg(\prod_{k=0}^{n-1} \cos(\phi_{n,k})\bigg)   \, G_n  \,
\left[\,  {\mathcal R}_n^{(r,{\rm s})}(t) +  {\mathcal  R}_n^{(r,{\rm a})}(t) \,\right]  \, ,
\ee
and where $r=1,\, 2$. The time correlations carried by the kernel $\dot{W}(\tau)$ are in the functions
\be   \label{eq:Rsplit}
\begin{array}{rcl}
{\mathcal R}_n^{(1,{\rm s})}(t) &=& {\displaystyle \cos(\varphi_n )\Big[\, 
\mathcal{U}_{n,0} (t) -\tan(\phi_{n,0}) \sum_{\ell=1}^{n-1} \tan(\phi_{n,\ell})\,\mathcal{U}_{n,\ell}(t) \, \Big]\, , }  \\[4mm]
{\mathcal R}_n^{(1,{\rm a})}(t) &=& {\displaystyle  p_0 \sin(\varphi_n)   \sum_{\ell=0}^{n-1} \tan(\phi_{n,\ell}) \,
\mathcal{U}_{n,\ell}(t) 
 \, ,  }    \\[4mm]
{\mathcal R}_n^{(2,{\rm s})}(t) &=&-\, {\displaystyle  \cos(\varphi_n)  \tan(\phi_{n,0}) \sum_{\ell=1}^{n}
\mathcal{V}_{n,\ell}(t)\,\xi_\ell   \, ,  }     \\[4mm]
{\mathcal R}_n^{(2,{\rm a})}(t) &=&{ \displaystyle  p_0 \sin(\varphi_n)  
\sum_{\ell=1}^{n} \mathcal{V}_{n,\ell}(t) \,\xi_\ell  \, .     }
\end{array}
\ee
The expression (\ref{eq:Qsplit1}) with (\ref{eq:Qsplit2}) and (\ref{eq:Rsplit}) 
holds for linear dissipation with arbitrary memory-friction.

\subsection{The Ohmic case}

Of particular importance is the case of Ohmic dissipation, which is the limit $s\to 1$ in the spectral density of Eq.~(\ref{eq:Jspec}). In the Ohmic universality limit $\omega_{\rm c}\to \infty$ we have
\be
\begin{array}{rcl}     \label{eq:strictlyohm}
W(\tau) &=& {\displaystyle  2K\ln\Big[\frac{\omega_{\rm c}}{\pi T}\sinh\big( \pi T |\tau| \big)\Big] 
+ i\,\pi K {\rm sign}(\tau) \, ,     }    \\[4mm]
\dot{W}''(\tau) &=& {\displaystyle  \frac{M q_0^2}{2} \gamma(\tau) = 2\pi K \delta(\tau) \, ,   } 
\end{array}
\ee
 where $K$ is a dimensionless damping strength, $K = M \gamma_1 q_0^2/(2\pi)$.

Because the kernel ${\dot W}''(\tau)$ is memoryless,
the sojourn-sojourn correlation ${\cal U}_{n,\ell}(t)$ is restricted to the nearest-neighbour term $\ell = n-1$ in which
the intermediate sojourn has length zero,
\be 
{\cal U}_{n,n-1} =-  2\pi K\, \delta(t_{2n}-t_{2n-1}) \, .
\ee
Observing that a blip of zero length does not interact with other blips, the $\xi$-summation in 
Eq.~(\ref{eq:Qsplit2}) cancels all sojourn-sojourn correlations except those of the first with the second.
Thus we have ${\cal U}_{n,n-1}= \delta_{n,1}\, {\cal U}_{1,0}$.
In addition, the odd term ${\mathcal R}_1^{(1,{\rm a})}(t)$ is zero because the single  
blip left over has zero length, thus readily 
\be 
\average{Q(t)}_1 = \frac{\pi K}{2}\, \delta^2 \,t \, .
\ee
Here $\delta = \Delta\,{e}^{-W_{\rm adia}/2}$ is a renormalized transition amplitude which takes into account the polarization cloud of the bath modes
in the adiabatic limit. We have $W_{\rm adia}= 2K \ln(\omega_{\rm c}/\Delta_{\rm r})$ and hence 
$\delta=\Delta_{\rm r} \equiv  \Delta (\Delta/\omega_{\rm c})^{K/(1-K)}$  in the regime 
$T\ll \Delta$, and   $W_{\rm adia} =2K \ln[\omega_{\rm c}/(2\pi T)]$  and hence 
$\delta = \Delta_{\rm r} (2\pi T/\Delta_{\rm r})^K$  in the regime
$T\gg \Delta$.

Consider next the temperature-dependent term $\average{Q(t)}_2$. We see from the imaginary part of the
expression (\ref{eq:strictlyohm}) that the blip-sojourn correlations $X_{j,k}$ are restricted to the 
nearest-neighbour term,
 $X_{j,k} =\pi K\, \delta_{j,k+1} $. Thus we have $\phi_{n,k} = \pi K \xi_{k+1}$. With this the expression
(\ref{eq:Qsplit2}) with (\ref{eq:Rsplit})  takes the form
\be  \label{eq:Q2ohm}
\begin{array}{rcl}
 {\average {Q(t)}}_2 &=& {\displaystyle \frac{1}{2}\, 
\sum_{n=1}^\infty  \Big( \frac{-\Delta^2}{2}\Big)^n  \cos(\pi K)^n
\int_0^t \mathcal{D}_{2n} \{t_j\} \sum_{\{ \xi_j=\pm 1\}}  \!     G_n     } \\[5mm]
&&{ \displaystyle  \quad \times \;
\Big(  -\cos(\varphi_n)\tan(\pi K)\, \xi_1    + p_0 \sin(\varphi_n) \Big)      
 \sum_{\ell=1}^n {\cal V}_{n,\ell} (t) \, \xi_\ell  } \,  .
\end{array}
\ee

At this point it is expedient to turn to  the average heat power  $\average { P(t)}$, 
which is the time derivative of the heat, $\average{ P(t)} = {\average {\dot Q(t)}}$. 
Since ${\cal V}_{n,\ell}(t=t_{2n})=0$, the  derivative of  ${\average {Q(t)}}_2$  is restricted to the derivative of ${\cal V}_{n,\ell}(t=t_{2n})$ in Eq.~(\ref{eq:Q2ohm}). Observing that $\ddot{W}'(\tau)$ coincides with $L'(\tau)$ given in Eq.~(\ref{eq:polcorr}), we readily obtain
\be  \label{eq:Q22ohm}
\begin{array}{rcl}
\fl {\average {P(t)}} &=& {\displaystyle  \frac{\pi K}{2} \delta^2 \, -\,    \frac{1}{2}\, 
\sum_{n=1}^\infty  \Big( \frac{-\Delta^2}{2}\Big)^n  \cos(\pi K)^n
\int_0^t \mathcal{D}_{2n} \{t_j\} \sum_{\{ \xi_j=\pm 1\}}  \!     G_n     } \\[5mm]
&\times&{ \displaystyle  
\Big( \cos(\varphi_n)\tan(\pi K)\, \xi_1    - p_0 \sin(\varphi_n) \Big)      
 \sum_{\ell=1}^n \big[\,L'(t-t_{2\ell-1})- L'(t-t_{2\ell}   )\,\big]\, \xi_\ell  } \, .
\end{array}
\ee
This form is the exact formal path sum expression for the average heat power of the two-state system
in the Ohmic scaling limit. Apart from the last factor, the expression represents the dynamics of the RDM of the TSS under all the time-correlations carried by the intra- and inter-blip correlation factor $G_n$. The residual factor with the time correlations of all  intermediate flip times  with the final time $t$ is specific to the mean of the heat power.
The expression (\ref{eq:Q22ohm}) may be used to calculate the time dependence of the heat power for any
 time-dependent driving of the system according to Eq.~(\ref{eq:bias}).

To sound out validity and potential of the presented method,  we now study 
the expression  (\ref{eq:Q22ohm}) for a constant bias $\epsilon_0^{}$. First, we see that in the term of $\Delta^{2n}$
the blip $\xi_\ell$, which may take any position in the sequence of $n$ blips, plays a particular role.
 In the weak damping limit, we may safely disregard  the interblip correlations
of the preceding and  the subsequent blips with the blip $\ell$,  $\Lambda_{j,\ell}=0$ and  $\Lambda_{\ell,k}=0$,  
as well as all interblip correlations across this blip, i.e., $\Lambda_{j,k}=0$,  where $j<\ell$ and $k>\ell$. Then
the  series (\ref{eq:Q22ohm}) can be expressed in terms of a convolution which involves in the  time segment the population correlation  function $\average{\sigma_z(\tau)}$~\cite{leggett1987dynamics}
and the coherence correlation  function $\average{\sigma_x(\tau)}$ of the TSS~\cite{grifoni1997coherences,grifoni1999dissipation,weiss1999quantum}. The resulting expression is
\be
\begin{array}{rcl} \label{eq:Pwd}
\fl\average{P(t)} &=& {\displaystyle \frac{\pi K}{2} \delta^2  - \frac{\Delta}{2} \left\{ \int_0^t d\tau\, 
\Big(\,\langle\sigma_x(t-\tau)\rangle_{\rm s}  \,\langle\sigma_z(\tau)\rangle_{\rm s} -
\langle\sigma_z(t-\tau)\rangle_{\rm a}\, \langle\sigma_x(\tau)\rangle_{\rm a} \right.   } \\[4mm]
&&
\qquad\;\; { \displaystyle -\, p_0 \,\big[
\langle\sigma_x(t-\tau)\rangle_{\rm a}  \,\langle\sigma_z(\tau)\rangle_{\rm s} -
\langle\sigma_z(t-\tau)\rangle_{\rm s}\, \langle\sigma_x(\tau)\rangle_{\rm a}   \, \big] \,\Big) \,L'(\tau)    } \\[4mm]
&&{\displaystyle \qquad \;\; +\,\left.   \tan(\pi  K)\, \int_0^t d\tau\, \big[\,  1- 
\langle\sigma_z(\tau)\rangle_{\rm s}  \,\big]\,L'(\tau)  \right\} \, . }
\end{array}
\ee
Here the components which are symmetric under  bias inversion $\epsilon_0^{}\to-\epsilon_0^{}$ have index s, and the respective antisymmetric components have index a. The initial conditions are 
$\average{\sigma_z(0)}_{\rm a}=\average{\sigma_x(0)}_{\rm s}  =\average{\sigma_x(0)}_{\rm a}=0 $
and $\average{\sigma_z(0)}_{\rm s}=1$.

In the further  analysis of the average heat power $\average{P(t)}$ we can thus rely on analytic results for the
nonequilibrium correlation functions $\langle\sigma_z(t)\rangle$ and $\langle\sigma_x(t)\rangle$ reported
in the recent literature~\cite{leggett1987dynamics,gorlich1989low,weiss1999quantum}.

To conclude this subsection, we consider as a preparatory work for the subsequent studies 
the total heat transferred from the TSS to the reservoir in the weak damping regime. Writing the
localized  states $\ket{ L}$ and  $\ket{ R}$  as linear combinations of the ground state 
$\ket{ g}$ and  excited state $\ket{ e}$ of $H_{\rm S}$, and assuming initial preparation according to Eq.~(\ref{eq:ini}) 
and thermal occupation of  the states $\ket{ g}$ and $\ket{ e}$ at time infinity, we obtain
\be  \label{eq:Qtotwd}
Q_{\infty}  \equiv \average{ Q(t\to\infty) }=\average{E}_{\rm ini} - \average{E}_{\rm eq} = 
\frac{p_0\,\epsilon_0^{}}{2} + \frac{\sqrt{\delta^2 + \epsilon_0^2}}{2} 
\tanh\Big( \frac{ \sqrt{\delta^2 + \epsilon_0^2} }{2T}\Big)  \, ~,
\ee
where $\average{E}_{{\rm ini}}$ and $\average{E}_{{\rm eq}}$ indicate the initial and equilibrium energy of the system respectively.
In the following subsections, we shall apply these results to study the heat transfer for weak damping 
in the Markov and in the quantum noise regime.

\subsection{Markov regime}

In the temperature regime $T\gg  \sqrt{\delta^2+\epsilon_0^2}$, 
the power spectrum $S(\omega)= K \pi\,\omega \coth(\frac{\omega}{2T})$  of the collective bath mode 
${\cal E}(t) =q_0\sum_\alpha c_\alpha x_\alpha (t)$ is essentially white,
\be 
S(\omega) =  \vartheta\, ,  \qquad\mbox{where}\qquad\vartheta \equiv 2 \pi K \,T \, .
\ee
Then the real parts of the polarization correlation function $L(\tau)$  and the flip-pair correlation function $W(\tau)$ of Eq.~(\ref{eq:strictlyohm}) take the form
\be\label{eq:Lmarkov}
L'(\tau) = \vartheta\, \delta(\tau)  \,  , \qquad\mbox{and}\qquad 
W'(\tau) = W_{\rm adia} + \vartheta\,|\tau| \,  ,
\ee
where $W_{\rm adia}= 2K \ln[\omega_{\rm c}/(2\pi T)]$ is the adiabatic contribution which dresses the
transfer matrix element according to $ \delta = \Delta\, {e}^{-W_{\rm adia}/2} =
\Delta_{\rm r} (2\pi T/\Delta_{\rm r})^{K}$.

Observing that the function $L'(\tau)$ is memory-less and the initial conditions for 
$\average{\sigma_z(t)}$ and $\average{\sigma_x(t)}$ in Eq.~(\ref{eq:Pwd})  are
$\average{\sigma_z(0)}=1$ and $\average{\sigma_x(0)}=0$,  the expression (\ref{eq:Pwd}) for the
heat power reduces to the concise form
\be \label{eq:Pwdmark1}
\average{P(t)} = \frac{\pi}{2} K \delta^2 - \pi K T\Delta \, [\,\langle\sigma_x(t)\rangle_{\rm s} -p_0 \langle\sigma_x(t)\rangle_{\rm a} \, ]  \; .
\ee
Since the pair interaction $W'(\tau)\propto \tau$, the interblip correlations $\Lambda_{j,k}$ are zero. As a result, the
series for the coherence correlation function $\average{\sigma_x(t)}$ is easily summed in Laplace space, yielding~\cite{grifoni1997coherences,weiss1999quantum}
\be\label{eq:sigxlapl}
\langle\hat\sigma_x(\lambda)\rangle = \frac{\delta^2}{\Delta\,  \lambda}\, 
\frac{K \pi [\delta^2 +\lambda(\vartheta +\lambda)]  +\, \epsilon_0^{}\lambda}{
\delta^2 ( \vartheta+\lambda) +\lambda[\,\epsilon_0^2+ (\vartheta+\lambda)^2\,] }  \, .
\ee
Importantly, the residuum of the pole at the origin, which is the equilibrium value
$\langle\sigma_x\rangle_{\rm eq}= \delta^2/(2T \Delta)$ of $\langle\sigma_x(t)\rangle$
reached at time infinity, together with the factor $\pi K T \Delta$ cancels exactly the constant term
$\pi K \delta^2/2$ in Eq.~(\ref{eq:Pwdmark1}). This secures that the heat power in fact vanishes as $t\to\infty$,
and the dynamics of $\average{P(t)}$ is determined by the simple zeros $\lambda_1$, $\lambda_2$ and $\lambda_3$ of the
denominator in Eq.~(\ref{eq:sigxlapl}).
The resulting expression is
\be \label{eq:Pmark}
\average{P(t)} =  \pi K T \delta^2 \, \left(   
\frac{p_0\, \epsilon_0\lambda_1- K \pi [\,\delta^2 +(\vartheta +\lambda_1) \lambda_1 \, ]}{
  \lambda_1 (\lambda_1-\lambda_2)(\lambda_1-\lambda_3)  } \-{e}^{\lambda_1 t}\,+ \,
\mbox{cycl.}
 \right)   \, ,
\ee
where "cycl." denotes addition of the terms obtained by cyclic permutation of the indices.
The behaviours  of $\lambda_1$, $\lambda_2$ and $\lambda_3$ reflect the characteristics of a cubic equation with real
coefficients, which depend on $\delta$, $\epsilon$ and the scaled temperature $\vartheta$.
 Physically, it is natural to write them in terms of an oscillation frequency $\Omega$, a decoherence rate
$\gamma$ and a relaxation rate $\gamma_{\rm r}$  as $\lambda_{1,2} = - \gamma \mp i\,\Omega$ and $\lambda_3 =-\gamma_{\rm r}$.

In the temperature regime $\sqrt{\delta^2+ \epsilon_0^2}\ll  T \ll \sqrt{\delta^2+  \epsilon_0^2}/(2\pi K)$ 
the bath coupling $H_{\rm I}$ is a perturbation. In the leading one-boson exchange  approximation, the rates increase
linearly with the scaled temperature $\vartheta$,
\begin{eqnarray}  \nonumber
\gamma_{\rm r} &=& \frac{\delta^2}{\delta^2+  \epsilon_0^2} \, \vartheta + {\cal O }(\vartheta^3)\,   \\
\label{eq:onephonon}
\gamma &=& \frac{\delta^2+2 \epsilon_0^2}{2(\delta^2+\epsilon_0^2)} \, \vartheta  +  {\cal O }(\vartheta^3)\,   \,   \\
\Omega &=& \sqrt{ \delta^2+  \epsilon_0^2} + {\cal O }(\vartheta^2)\,  .   \nonumber
\end{eqnarray}

As temperature is raised, multi-boson exchange processes become increasingly important. In the
temperature range  well above $\sqrt{\delta^2+ \epsilon_0^2}/(2\pi K)$,
the decoherence and relaxation rates show drastically different behaviours,
\begin{eqnarray} \nonumber
\gamma_{\rm r} &=& \frac{\delta^2}{\vartheta} + {\cal O }(1/\vartheta^3)\,   \\ \label{eq:multiphonon}
\gamma &=&  \vartheta - \frac{\delta^2}{\vartheta}   +  {\cal O }(1/\vartheta^3)\,   \,   \\   \nonumber
\Omega &=&  \epsilon_0 +\frac{\delta^2 (4\epsilon_0^2 -\delta^2)}{8\epsilon_0^3}\,\frac{1}{\vartheta^2}
+ {\cal O }(1/\vartheta^4)\,     \,  .
\end{eqnarray}
In this so-called Kondo regime, the relaxation rate decreases with increasing temperature as $T^{2K-1}$, whereas
the decoherence rate increases linearly with $T$.

Integration of the expression  (\ref{eq:Pmark}) yields the average heat $\average{Q(t)}$ transferred to the reservoir until time $t$. 
Readily the average heat transferred until time infinity is found from Eq.~(\ref{eq:Pmark})  in the Markovian regime as
\be\label{eq:Qinfmark}
Q_{\infty} = \frac{p_0 \,\epsilon_0^{}}{2} + \frac{\delta^2
+\epsilon_0^2}{4T} \,  .
\ee
{ It is straightforward to see by  use of the Vieta relations for the frequencies $\lambda_1$, $\lambda_2$ and $\lambda_3$, that the expression  (\ref{eq:Qinfmark}) holds for all $T$ in the Markov regime.
Importantly, the result (\ref{eq:Qinfmark}) is just the high-temperature limit of the expression (\ref{eq:Qtotwd}).}

\subsection{Quantum noise regime}

\begin{figure}
\begin{minipage}[b]{0.45\linewidth}
\begin{overpic}[width=\linewidth]{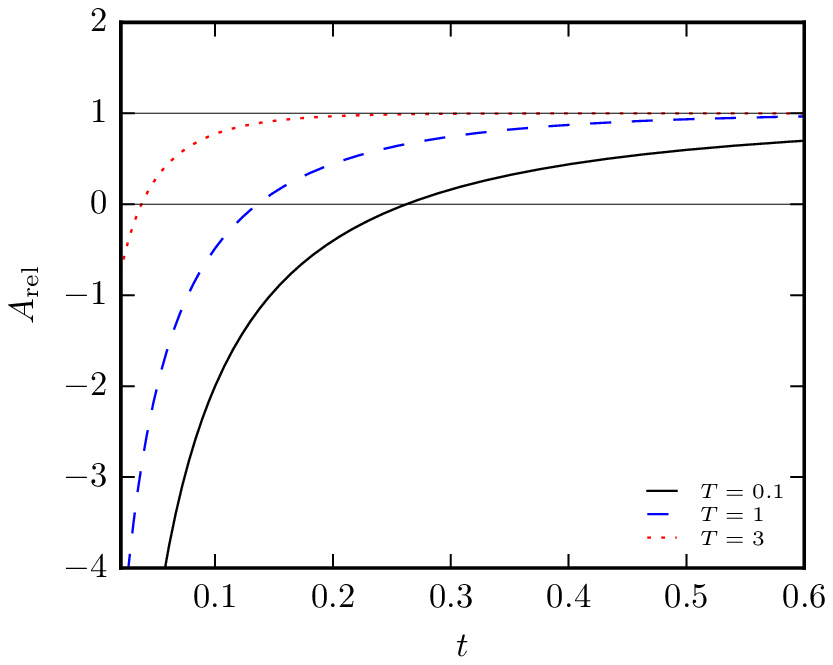}\put(2,60){a)}\end{overpic}
\end{minipage}
\quad
\begin{minipage}[b]{0.45\linewidth}
\begin{overpic}[width=\linewidth]{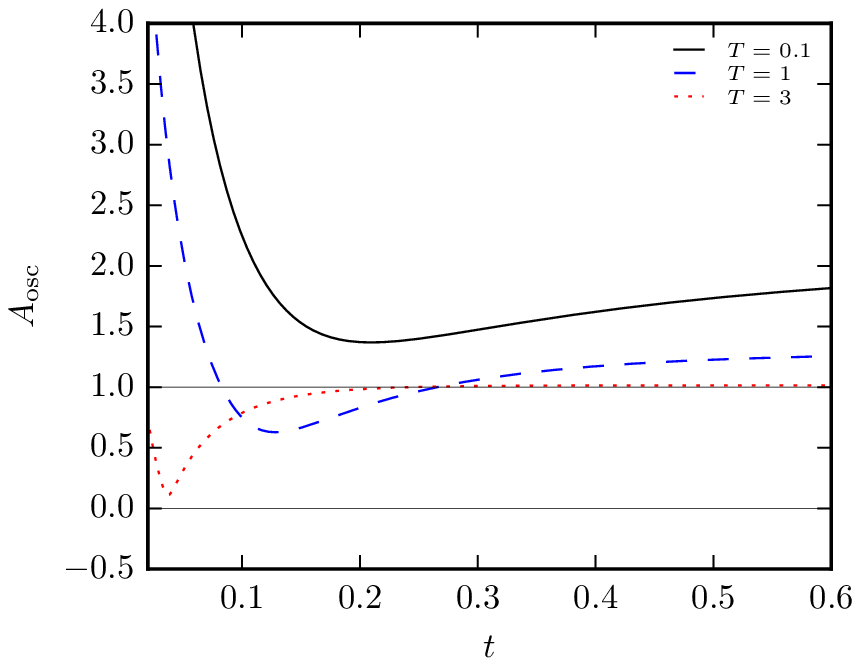}\put(2,60){b)}\end{overpic}
\end{minipage}
\caption{{The amplitudes $A_{\rm rel}$ and $A_{\rm osc}$ are plotted versus time for different temperatures,
$T = 0.1$ (black solid), $T= 1.$ (blue dashed), and $T = 3.$ (red dotted). The parameters are
$\delta=\epsilon_0=1.$, $\omega_{\rm c}=50.$, and $K=0.02$. Times, frequencies and temperature are 
scaled with $\delta$. The time  where $A_{\rm rel}$ changes sign
and the time where it approaches unity move towards the origin as temperature is increased.
Correspondingly, the minimum of $A_{\rm osc}$ moves to the origin, as $T$ is increased.  Other characteristics of
$A_{\rm rel}$ and $A_{\rm osc}$ are discussed in the main text.   }

\label{fig:amplitudes}}
\end{figure}

\begin{figure}
\centering
\includegraphics[scale=0.9]{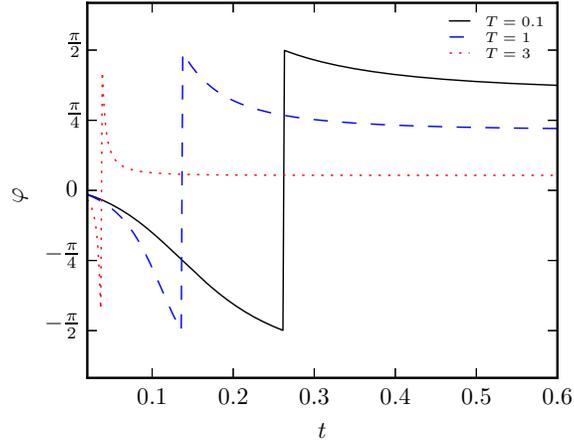}
\caption{The phase $\varphi$ is plotted as a function of time for different temperatures, 
$T = 0.1$ (black solid), $T= 1.$ (blue dashed), and
$T = 3.$ (red dotted). The other parameters are the same as in Fig.~\ref{fig:amplitudes}. 
The phase jump by $\pi$ occurs
at the instant where the amplitude $A_{\rm rel}$ changes sign. The constant value at asymptotic times 
increases with decreasing temperature. 
}
\label{fig:phase}
\end{figure}

\begin{figure}
\begin{minipage}[b]{0.45\linewidth}
\begin{overpic}[width=\linewidth]{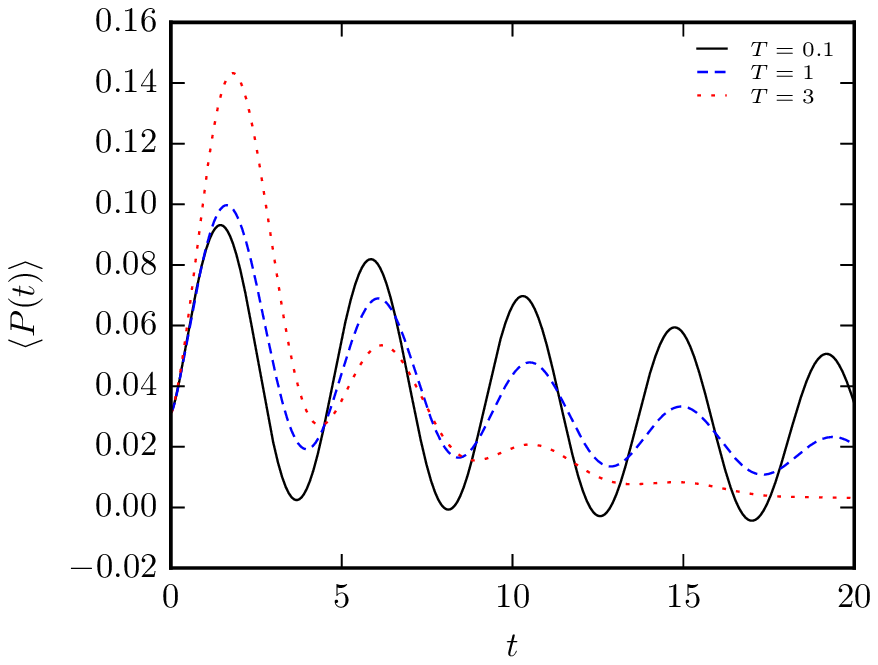}\put(2,60){a)}\end{overpic}
\end{minipage}
\quad
\begin{minipage}[b]{0.45\linewidth}
\begin{overpic}[width=\linewidth]{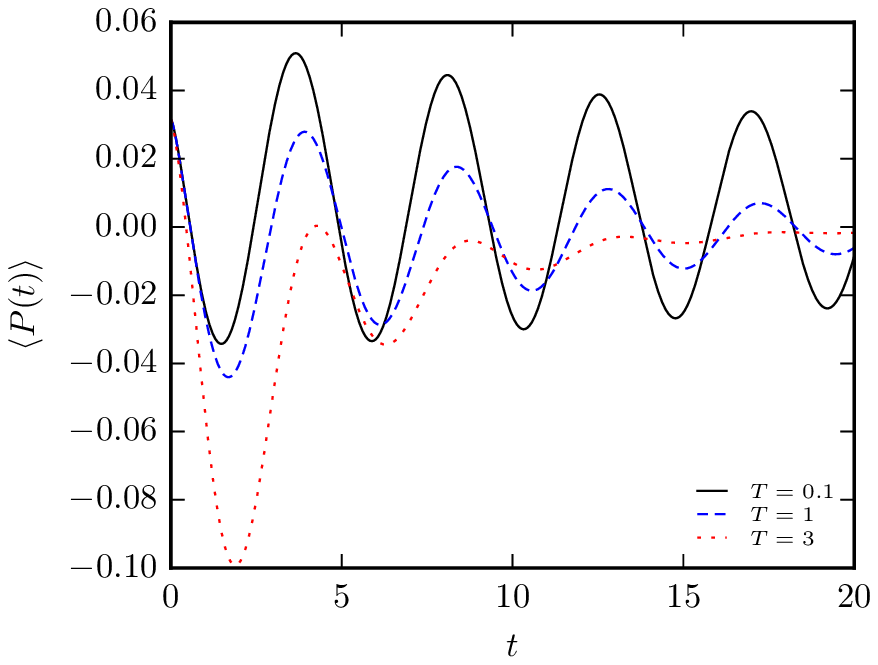}\put(2,60){b)}\end{overpic}
\end{minipage}
\caption{{The mean heat power $\average{P(t)}$ (in units of $\delta^2$) is plotted as a function of time for
 different temperatures, $T = 0.1$ (black solid), $T= 1.$ (blue dashed),  and $T = 3.$ (red dotted). The parameters are
$\delta=\epsilon_0=1.$, $\omega_{\rm c}=50.$, and $K=0.02$. 
Times, frequencies and temperature are scaled with $\delta$. The initial condition is $p_0=1$ in panel a) and
$p_0=-1$ in panel b). The phases of the oscillations in panel b) differ from the corresponding ones in panel a) by $\pi$.
The quality of the oscillations decreases with increasing $T$.   }

\label{fig:power}}
\end{figure}

\begin{figure}
\begin{minipage}[b]{0.45\linewidth}
\begin{overpic}[width=\linewidth]{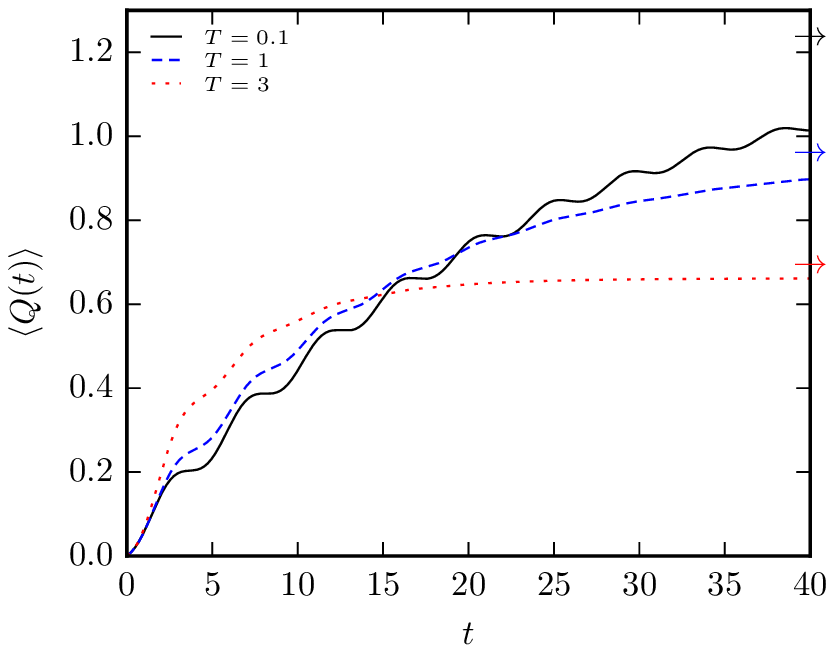}\put(2,60){a)}\end{overpic}
\end{minipage}
\quad
\begin{minipage}[b]{0.45\linewidth}
\begin{overpic}[width=\linewidth]{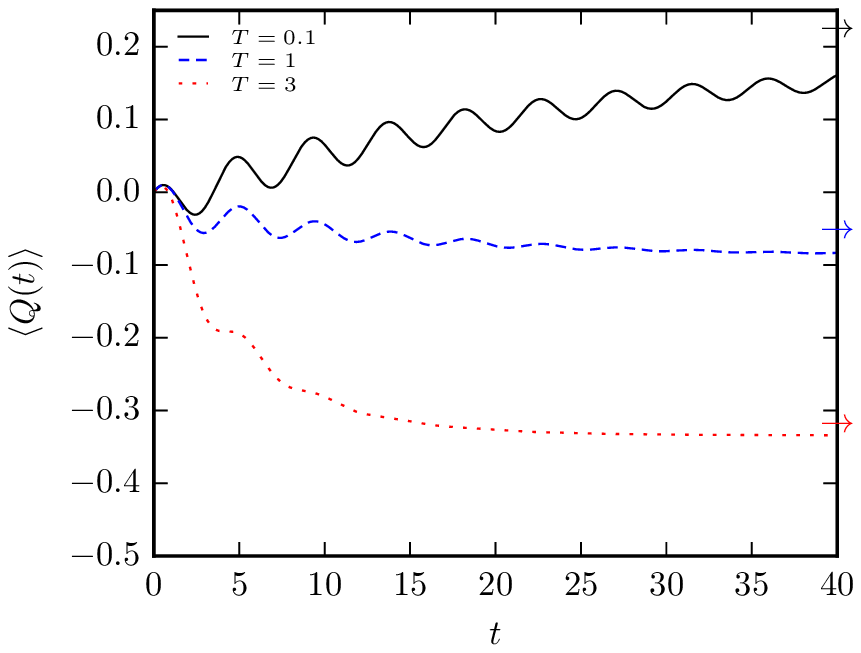}\put(2,60){b)}\end{overpic}
\end{minipage}
\caption{{The mean heat $\average{Q(t)}$ (in units of $\delta$) is plotted as a function of time for different temperatures,  $T = 0.1$ (black solid), $T= 1.$ (blue dashed),  and $T = 3.$ (red dotted). The parameters are
$\delta=\epsilon_0=1.$, $\omega_{\rm c}=50.$, and $K=0.02$. Times, frequencies and temperature are scaled with $\delta$. The initial condition is $p_0=1$ in panel a) and
$p_0=-1$ in panel b). 
The phases of the oscillations in panel b) differ from the corresponding ones in panel a) by $\pi$.
The quality of the oscillations decreases with increasing $T$. The respective total heat transferred, which is calculated from the expression (\ref{eq:Qtotwd}), is indicated by the arrow at the right border.  }

\label{fig:heat}}
\end{figure}

For temperatures of the order $\sqrt{\delta^2+\epsilon_0^2}$ or lower, the power spectrum $S(\omega)$ is coloured
because of quantum mechanical noise and the functions $L'(\tau)$ and $W'(\tau)$   carry memory. 
Nevertheless, it is possible to calculate exactly the one-boson contribution to the rates $\gamma$ and $\gamma_{\rm r}$~\cite{gorlich1989low,weiss1999quantum}. 
The resulting expressions are
\be \label{eq:quantrates}
\gamma_{\rm r}  =  \pi K \frac{\delta^2}{\Omega}\coth\Big( \frac{\Omega}{2T}\Big)  \, , \qquad \mbox{and}
\qquad \gamma = \frac{1}{2} \gamma_{\rm r} + 2\pi K T\frac{\epsilon_0^2}{\Omega^2}  \, ,
\ee
where $\Omega =\sqrt{\delta^2+\epsilon_0^2}$ and $\delta=\Delta_{\rm r}$.
The correlation functions $\average{\sigma_z(\tau)}$ and $\average{\sigma_x(\tau)}$ 
going down into the expression of Eq.~(\ref{eq:Pwd})  take the form
\be 
\begin{array}{rcl}
\average{\sigma_z(\tau)} &=& {\displaystyle  \frac{\epsilon_0^{}}{\Omega}\tanh\Big( \frac{\Omega}{2T}\Big) 
\Big[  1 -\, {e}^{-\gamma_{\rm r} \tau}  \Big] + \frac{\epsilon_0^2}{\Omega^2}\, {e}^{-\gamma_{\rm r} \tau}
+  \frac{\delta^2}{\Omega^2}\,\cos(\Omega \tau)\,   {e}^{-\gamma \tau}  } \, \\[4mm]
\average{\sigma_x(\tau)} &=& {\displaystyle  \frac{\delta^2}{\Delta \Omega}\tanh\Big( \frac{\Omega}{2T}\Big) 
\Big[  1 -\, {e}^{-\gamma_{\rm r} \tau}  \Big] + \frac{\epsilon_0^{} \delta^2}{\Delta\Omega^2}\, \Big[\,
{e}^{-\gamma_{\rm r} \tau}
- \cos(\Omega \tau)\,   {e}^{-\gamma \tau} \,\Big] }  \, .
\end{array}
\ee 
Next, the convolution in Eq.~(\ref{eq:Pwd}) is conveniently evaluated by using for $L'(\tau)$  
its spectral representation in Eq.~(\ref{eq:polcorr}) and writing the $\coth(\frac{\omega}{2T})$-function in terms of the  sum representation
\be
\coth\Big(
 \frac{\omega}{2T}\Big) = \sum_{n=-\infty}^{\infty} \frac{2 \omega T }{\omega^2 +\omega_n^2} \; ,
\ee
where $\omega_n =2\pi T \,n$ is a bosonic  Matsubara frequency.
Then the $\tau$-integration is straightforward.  The remaining  $\omega$-integral picks up (i) a contribution from
the singularity at $\omega =\Omega$ formed in the limit $K\to 0$ and (ii) a contribution from the residua of the  
infinite sequence of poles stringed along the imaginary axis at $\omega= i \, \omega_n$. This split-up into a system and a Matsubara part is a general feature of  equilibrium correlation functions in the quantum regime.

The  first contribution  has  a part which cancels the  first term in Eq.~(\ref{eq:Pwd}), and the remaining part is
\be \label{eq:Pqu1}
\average{P(t)}_1 =\frac{\pi}{2} K \delta^2 \, \left(  {e}^{-\gamma_{\rm r} t} + \frac{p_0\, \epsilon_0^{}}{\Omega}  
\coth\Big( \frac{\Omega}{2T}\Big) \big[\,  {e}^{-\gamma_{\rm r} t} -\cos(\Omega t)\, 
 {e}^{-\gamma  t}\,\big] \right)  \, .
\ee
The contribution from the Matsubara poles is odd in the bias and is
\be
\average{P(t)}_2 =   K  \delta^2\,  \frac{p_0\,\epsilon_0^{}}{\Omega} \Big(      
u_2(t)\,\big[\, \cos(\Omega t)\, {e}^{-\gamma  t} -  {e}^{-\gamma_{\rm r} t} \,\big]
 + u_1(t) \,\sin(\Omega t) \, {e}^{-\gamma  t}\,\big]    \Big)   \, . 
\ee
The time-dependent coefficients $u_1(t)$ and  $u_2(t)$ are 
\be \label{eq:Pqu2}
\begin{array}{rcl}
u_1(t) &=&{\displaystyle  2\pi T \sum_{n=1}^\infty \, \frac{\omega_n}{ \omega_n^2+\Omega^2}  
\, \big[\,   {e}^{-\omega_n/\omega_{\rm c} }     -  {e}^{-\omega_n(t+\frac{1}{\omega_{\rm c} } ) }  \,\big] } \, ,\\[5mm]
u_2(t) &=& {\displaystyle \frac{2\pi T }{\Omega}\,\sum_{n=1}^\infty \, \frac{ \omega_n^2}{ \omega_n^2+\Omega^2}
\,{e}^{-\omega_n(t+\frac{1}{\omega_{\rm c} })   } }  \, .
\end{array}
\ee
These functions can be written as linear combinations of hypergeometric $_{2}F_1$-functions which reduce to
sine and cosine integral functions at $T=0$. The function $u_1(t)$ depends logarithmically on the cutoff
$\omega_{\rm c}$.

As  the Markov regime at $T$  well above $\Omega$ is approached, 
the expression (\ref{eq:Pqu1}) smoothly matches on the expression (\ref{eq:Pmark}) with the one-boson rates (\ref{eq:onephonon}). The heat power term $\average{P(t)}_2$ is a pure quantum contribution, which is negligibly small in this regime. Conversely, as the temperature is lowered, quantum effects appear in 
$\average{P(t)}_1$ through the $\coth$-function in the amplitude and in the rate
expressions (\ref{eq:quantrates}). This function indicates emission and absorption of quanta with energy $\omega$
in thermal equilibrium. Moreover, the term $\average{P(t)}_2$  becomes increasingly important as $T$ is decreased,
and may even dominate the time-dependence  of the heat power at low $T$. The combined expression can be conveniently written in terms of the amplitudes
\be
\fl \qquad A_{\rm rel} (t) =1 - \frac{2}{\pi} \tanh\Big(  \frac{\Omega}{2T} \Big) \,u_2(t) \, ,\qquad
B(t) =  \frac{2}{\pi} \tanh\Big(  \frac{\Omega}{2T} \Big) \, u_1(t)    \, .
\ee
With these forms, the antisymmetric part of the total  average heat power 
$\average{P(t} = \average{P(t)}_1+\average{P(t)}_2$ is conveniently written in terms of  the amplitude
$A_{\rm rel}(t)$ of the relaxation contribution, the amplitude $A_{\rm osc}(t)$ of the oscillatory contribution, and a phase shift $\varphi(t)$,
\be
A_{\rm osc}(t) =   \sqrt{A_{\rm rel}(t)^2 + B(t)^2} \, ,\qquad\quad
\varphi(t) = \arctan\Big(\frac{B(t)}{A_{\rm rel}(t) }  \Big)  \, .    \\[4mm]
\ee
The resulting expression for the total mean heat power is
\beq  \label{eq:heatfinal}
\fl
\average{P(t)} &=& \average{P(t)}_1 + \average{P(t)}_2 = \nonumber \\
\fl &=&  \frac{\pi}{2} K \delta^2 \left\{  \,   {e}^{-\gamma_{\rm r} t} 
+  \,\frac{p_0\,\epsilon_0^{}}{\Omega} \coth\Big( \frac{\Omega}{2T}  \Big)
\Big( A_{\rm rel}(t)  \, {e}^{-\gamma_{\rm r}  t} -A_{\rm osc}(t)\,{\rm sign}[\varphi(t)]\,
\cos[\,\Omega t +\varphi(t)\,]   
\, {e}^{-\gamma  t}             \Big)     \right\}    \, .  
\eeq
The oscillatory part of $\average{P(t)}$, and equivalently of the integrated power $\average{Q(t)}$, 
describes seesaw transport of heat between system and reservoir.
The damped oscillatory parts are superimposed by relaxation contributions.

Considering Eq.~(\ref{eq:heatfinal}) and comparing it with Eq.~(\ref{eq:Pqu1}), one can deduce that the degree of deviation from unity  of the amplitude factors $A_{\rm rel}(t)$ and $A_{\rm osc} (t)$
 indicate the relative strength of the pure quantum contribution
$\average{P(t)}_2$. Indeed, in the case $A_{{\rm rel}}(t)=A_{{\rm osc}}(t)=1$ and $\varphi(t)=0$ one recovers the form (\ref{eq:Pqu1}) in which the  pure quantum contribution $\average{P(t)}_2=0$ is absent. The amplitudes $A_{\rm rel}(t)$ and $A_{\rm osc} (t)$, and the phase
$\varphi(t)$  are plotted in Figs.~\ref{fig:amplitudes} and \ref{fig:phase} as a functions of time for  three different temperatures. 
The amplitude $A_{\rm rel}(t)$ approaches  unity in the regime $ t>1/\Omega$ for $T\ll\Omega$ and in the regime
$t>1/T$ for $T\gg\Omega$.
For fixed $T$, the amplitude  $A_{\rm rel}(t)$  is monotonically decreasing as time $t$ is lowered and 
eventually changes sign at time $t_0$. In the temperature regime $T\gg\Omega$, the instant $t_0$  
depends inversely on $T$ as $t_0\approx \ln(2)/(2\pi T) $ and eventually enters
 the core region, $t_0= {\cal O}(1/\omega_{\rm c}),$ at temperatures of the order $\omega_{\rm c}$. 
Reversely, the instant $t_0$ increases with
decreasing temperature and reaches at $T=0$ the maximal value $t_0=1/(\pi \Omega)$.

The amplitude $A_{\rm osc}(t)$ has a minimum at the time $t_0$, at which
$A_{\rm rel}(t)$ is zero, whereas the phase $\varphi(t)$ undergoes a sudden jump 
from $-\pi/2$  to $\pi/2$ as t passes $t_{0}$ from below.
At times  below $t_0$, the  pure quantum contribution $\average{P(t)}_2$ dominates the behaviour of the heat power $\average{P(t)}$.  Both the amplitudes$A_{\rm osc}(t)$ and the phase $\varphi(t)$ approach constant values $A_{{\rm osc,}\infty}$ and $\varphi_\infty$ at times $t\gg1/\Omega$. 
These values get larger as $T$ is decreased, and  at low $T$ they are 
notably different from the values $A_{\rm osc}=1$ and $\varphi=0$ holding in the absence of the pure quantum contribution $\average{P(t)}_2$.
The asymptotic values are largest at $T=0$ and are
\be
A_{{\rm osc},\infty}  =  \sqrt{1 +B_\infty^2} \, , \qquad\mbox{and}\qquad \varphi_\infty = \arctan(B_\infty)   \, ,
\ee
where $B_\infty =  \arctan[2\ln(\omega_{\rm c}/\Omega) - C_{\rm E}\,]$, and where $C_{\rm E}$ is Euler's constant.
This shows that the direct time dependence of $\average{P(t)}$ and $\average{Q(t)}$at low $T$ 
strongly depends on the
pure quantum contribution $\average{P(t)}_2$ for all times. These non-Markovian effects could be measured at temperature $T< \Omega$, provided that the detector used can resolve the signal with the sufficient accuracy for time $t < 1/\gamma, 1/\gamma_{{\rm r}}$.

On the bottom line,  the contribution of $\average{P(t)}_2$ to the overall heat
$Q_\infty$ is negligibly small. In the asymptotic weak-damping limit, the total heat transferred to the bath is
found from the expression (\ref{eq:heatfinal}) as
\be
Q_\infty = \frac{\pi}{2} K\Delta^2 \Big[\, 1 + p_0 \frac{\epsilon_0^{}}{\Omega}  
\coth\Big(  \frac{\Omega}{2T} \Big) \,\Big]\, \frac{1}{\gamma_{\rm r} }     \, .
\ee
With the form (\ref{eq:quantrates}) for $\gamma_{\rm r}$, $Q_{\infty}$ coincides with the expression (\ref{eq:Qtotwd}).

Plots of $\average{P(t)}$ and $\average{Q(t)}$ are shown for different temperatures and different initial conditions
in Figs.~\ref{fig:power} and \ref{fig:heat}.
The oscillations of $\average{Q(t)}$ are clearly visible for the case $T=0.1\; \delta$ both for initial condition
$p_0=1 $ in panel a) and for $p_0=-1$ in panel b). For $p_0=1$, the three curves stay together until time 
$t\approx 15\; \delta$ and then diverge  to reach the respective asymptotic values indicated by the arrows on the right border. We see from the curves in Fig.~\ref{fig:heat} b), which correspond to the initial condition $p_0=-1$,
that the heat can be positive or negative, dependent on temperature.

\section{Conclusions}
\label{sect:concl}

In this paper we established the functional integral description for the time-dependent heat exchange
of a quantum system coupled to a thermal reservoir. We presented the exact formal solution for the 
moment generating functional  which carries all statistical features of the heat exchange process for general linear dissipation.
We derived an exact formal expression for the transferred heat and applied this formalism to the dissipative two-state system.
We showed that the difference between the  dynamics of the heat transfer and  the dynamics of the reduced density matrix (RDM) is an additional time-nonlocal correlation function which correlates intermediate states of the RDM 
with the final state.

To investigate the potential of the presented method,
we calculated the dynamics of the average heat power and average heat in analytic form for weak Ohmic dissipation both in the Markovian regime relevant at high temperatures and in the non-Markovian quantum noise regime holding 
when temperature is of the order of the level splitting $\Omega$ or lower.
 In the latter regime, the heat is represented by a convolution integral
which involves the  population and coherence correlation functions of the dissipative TSS and the polarization correlation function of the reservoir. We find that the heat transfer receives contributions both from singularities 
related to the system dynamics  and from Matsubara singularities resulting from the system-reservoir correlations.
The latter yield significant contributions in the quantum noise regime while they are absent in the Markovian regime.

Altogether we have achieved a complete description of the dynamics of the heat transfer for weak damping
ranging from the classical regime down to zero temperature

\appendix
\section{Gaussian integration}
\label{app:gaussianintegration}

In this Appendix we outline  the evaluation of the multiple integral expression  (\ref{eq:influence}).
For simplicity, we denote  quantities related to the reservoir oscillator $\alpha$ by an index $\alpha$, 
and drop the redundant  label ${\rm R}$. Further, we use the inverse temperature $\beta =1/T$, but
return to $T$ when we employ the expressions (\ref{eq:kernels_l}) in Section \ref{sect:path}.

At first, we see that the influence functional $\mathcal{F}^{(\nu )}[q,q']$ can be written in the product form
\be
\label{eq:mfl_alpha}
\mathcal{F}^{(\nu )}[q,q']=\prod_{\alpha=1}^N \mathcal{F}_\alpha^{(\nu)}[q,q'] \, ,
\ee
in which the term $\mathcal{F}_\alpha^{(\nu)}[q,q']$ is the contribution of the bath oscillator $\alpha$. 
We have
\begin{equation}
\fl  \mathcal{F}^{(\nu)}_\alpha [q,q']= \int dx_{\alpha f} dx'_{\alpha f} dx_{\alpha i} dx'_{\alpha i} 
\matrixel{ x_{\alpha i}}{{\rho^{(\nu)}}_{\alpha}(0)}{x'_{\alpha i}}
 \matrixel{x'_{\alpha f}}{e^{i \nu H_{\alpha}}}{ x_{\alpha f}} F[q, x_{\alpha f}, x_{\alpha i}] 
F^*[q', x'_{\alpha f}, x'_{\alpha i}]  \, .
\label{eq:inflalpha}
 \end{equation}
The matrices of the density operators
$\rho^{(\nu)}_\alpha (0) = {e}^{-(\beta +i\,\nu)H_\alpha}/Z_\alpha$ and $ {e}^{i\,\nu H_\alpha}$
depend solely on  coordinates of the bath oscillator $\alpha$ and may be written as
\begin{equation}
\begin{array}{rcl}
  \matrixel{x_{i\alpha}}{{\rho}^{(\nu)}_{\alpha} (0)}{x'_{i \alpha}} &=& {\displaystyle \frac{1}{Z_{\alpha}} 
 {\cal A}_{1 \alpha} (\beta +i \nu)
 \exp{ \big\{- {\cal B}_{1 \alpha} (x_{i \alpha},x'_{i \alpha};\beta + i\, \nu)\big\} }   \, ,    }\\[4mm]
\matrixel{x'_{f \alpha}}{e^{i\nu H_{\alpha}}}{x_{f \alpha}} &=&  {\displaystyle
 {\cal A}_{1 \alpha} (-i \nu )
 \exp \Big\{ -{\cal B}_{1 \alpha} (x_{f \alpha},x'_{f \alpha};-i\, \nu) \Big \}   \, .   }
\end{array}
\end{equation}
The prefactor ${\cal A}_{1 \alpha} (\tau) $ and the Euclidean action $ {\cal B}_{1 \alpha} (x,y;\tau)$ 
for an imaginary-time interval $\tau$ are given by
\be \label{eq:abfunctions}
\begin{array}{rcl}
  {\cal A}_{1 \alpha} (\tau) &=&  {\displaystyle\Big[\frac{m_\alpha \omega_\alpha}{2\pi \sinh(\omega_\alpha\tau )}
\Big]^{1/2} \, ,   }   \\[4mm]

 {\cal B}_{1 \alpha} (x,y;\tau)  &=& {\displaystyle  \frac{m_\alpha \omega_\alpha}{2 \sinh (\omega_\alpha \tau)}
 [(x^2 +  y^2)\cosh ( \omega_\alpha \tau) - 2 x y ]   \, , }
\end{array}
 \ee
and $Z_\alpha$ is the partition function of the reservoir mode $\alpha$,
\begin{equation}
 Z_\alpha= \frac{1} {2 \sinh(\beta \omega_\alpha/2)}   \, .
\end{equation}

Correspondingly, the propagator of the driven oscillator $\alpha$  takes the form
\be
 F_\alpha [q, x_{f \alpha}, x_{i \alpha}] = {\cal A}_{2 \alpha} ( t)
 \exp{ \big\{i \phi_\alpha[q, x_{f \alpha}, x_{i \alpha};t]\big\} }
\label{eq:fampl}
\ee
with the prefactor
\be
 {\cal A}_{2 \alpha} (t) = \Big[\frac{m_\alpha \omega_\alpha}{2\pi \sin(\omega_\alpha t )} \Big]^{1/2} \, .
\ee
The phase is determined by the real-time action in the presence of the external force 
$c_\alpha q(\tau)$~\cite{feynman1963theory,weiss1999quantum}
\be
  \fl \phi_\alpha[q, x_{f \alpha}, x_{i \alpha};t] =  {\cal B}_{2 \alpha} (x_{i \alpha}, x_{f \alpha},  t)   
+x_{i \alpha}\int_0^t dt' \,{\cal C}_\alpha ( t-t')q(t') + 
    x_{f \alpha}\int_0^t\! \! dt' \,{\cal C}_\alpha (t')q(t') +\psi_\alpha[q] \, ,  \label{eq:minkaction}
\ee
where
\be     \label{eq:abfunc} 
\begin{array}{rcl}
 {\cal B}_{2 \alpha} (x,y; t)  &=& {\displaystyle \frac{m_\alpha \omega_\alpha}{2 \sin(\omega_\alpha t)}
 [(x^2 +  y^2)\cos ( \omega_\alpha t ) - 2 x y ]   \, ,   }  \\[4mm]
 {\cal C}_\alpha(t') &=&{ \displaystyle \frac{c_\alpha}{\sin(\omega_\alpha t)} \sin(\omega_\alpha t')   \, .}
\end{array} 
\ee
The first term in Eq.~(\ref{eq:minkaction}) is the phase  resulting from the internal oscillator dynamics, the second and third terms depend linearly on the initial and final oscillator position and on the history of the system. Finally, $\psi_\alpha[q]$ is a global phase
which does not depend  on the  reservoir mode,
\be
\psi_\alpha[q] = -\frac{\mu_\alpha}{2}\int_0^t \! \! dt'q^2(t') - \frac{\sin(\omega_\alpha t)}{m_\alpha \omega_\alpha} \int_0^t dt' \int_0^{t'}dt'' {\cal C}_\alpha (t-t') {\cal C}_\alpha (t'') q(t') q(t'')  \, ,
\ee
where $\mu_\alpha = c^2 _\alpha/(m_\alpha \omega^2 _\alpha)$. Here the first term originates from the counter
term in the interaction $H_{\rm I}$ given in  Eq.~(\ref{eq:Hint}).
We can thus write the generalized influence functional $\mathcal{F}^{(\nu)}[q,q']$ in the  form
\be
 \mathcal{F}_\alpha^{(\nu)} [q,q'] = \frac{1}{Z_{\alpha} } {\cal A}_{1\alpha} (\beta+i\,\nu) {\cal A}_{1\alpha} (-i\,\nu)
{\cal A}_{2\alpha}  (t)   {\cal A}_{2\alpha}^\ast  (t)   \,{e}^{ i \psi_\alpha[q]} \,  {\cal K}_\alpha [q,q'] ~,
\label{eq:math_F}
\ee
where $ {\cal K}_\alpha$ captures the integrations of the bath mode $\alpha$,
\begin{eqnarray}
\fl {\cal K}_\alpha&=& \int dx_{f \alpha} dx'_{f \alpha} dx_{i \alpha} dx'_{i \alpha} 
\exp{    \Big\{ i\, [
{\cal B}_{2 \alpha} (x_{i \alpha}, x_{f \alpha};  t) -
{\cal B}_{2 \alpha} (x'_{i \alpha},x'_{f \alpha}; t)] \Big\}  }    \nonumber   \\
\label{kkalpha}
\fl &&\times \;\;
\exp{
 \Big\{ - [
{\cal B}_{1 \alpha} (x_{i \alpha}, x'_{i \alpha}; \beta+i\,\nu)+ 
{\cal B}_{1 \alpha} (x_{f \alpha},x'_{f \alpha};-i\,\nu )
]
\Big\}  }    \\
\fl &&\times\;\; \exp{ \Big\{i\int_0^t dt' \big[  x_{i \alpha}{\cal C}_\alpha ( t-t')+ x_{f \alpha}{\cal C}_\alpha (t')]q(t') -[x'_{i \alpha}{\cal C}_\alpha (t-t')-x'_{f \alpha}{\cal C}_\alpha (t')] q'(t')\big] \Big \} }~.     \nonumber
\end{eqnarray}
The exponent of the integrand of the fourfold integral (\ref{kkalpha})  is a Gaussian form. Written in terms of the
symmetric and antisymmetric cordinates $x_{i\alpha}^{(\pm)} =(x_{i,\alpha}\pm x_{i\alpha}')/\sqrt{2}$ and
$x_{f\alpha}^{(\pm)} =(x_{f\alpha}\pm x_{f\alpha}')/\sqrt{2}$, the Gaussian form is free of mixed terms 
$x_{i\alpha}^{(+)} x_{f\alpha}^{(+)}$. Hence both the $x_{i\alpha}^{(+)}$- and the $x_{f\alpha}^{(+)}$-integrations may be performed by completing the square,
\begin{equation}
 \int_{-\infty}^\infty  dx\, e^{-a x^2 + b x}  = \sqrt{\pi/a}\;  e^{\frac{b^2}{4 a}} \, .
\label{gaussian}
\end{equation}
With the shift $x_{f\alpha}^{(-)} =y_{f\alpha}^{(-)} + x_{i\alpha}^{(-)} \cos(\omega_\alpha t)$ the arising quadratic form
of the exponent is free of the mixed term $y_{f\alpha}^{(-)} x_{i\alpha}^{(-)} $. 
Thus the remaining $y_{f\alpha}^{(-)} $-
and $ x_{i\alpha}^{(-)} $-integrals can be done by using again twice the relation (\ref{gaussian}).
The various pre-exponential factors that accrued  multiply to unity. In addition, the arising exponent is conveniently expressed in terms of the paths $\eta(\tau)$ and $\xi(\tau)$ given in Eq.~(\ref{etaxipaths}). After all,
the resulting expression for the influence functional $(\ref{eq:inflalpha})$ may be written in the form
\be  \label{eq:Finflapp}
 \mathcal{F}^{(\nu)}_\alpha [\eta,\xi] = {e}^{-\Phi_\alpha^{(0)} [\eta,\xi] } \;
{e}^{i\,\Delta\Phi_\alpha^{(\nu)}[\eta,\xi] }
\ee
with
\begin{eqnarray}
\hspace{-2.6cm} 
\fl &&\Phi^{(0)}_\alpha[\eta, \xi] = i\, \frac{\mu_\alpha q_0^2}{2} \int_0^t dt'  \eta(t')\xi(t') +
\int_0^t dt' \int_0^{t'} dt''  \xi(t') [ L_\alpha'(t'-t'') \xi(t'') + i L_\alpha'' (t'-t'') \eta(t'') ]
\label{eq:phizero}  \, ,\\
\hspace{-2.6cm} \fl &&\Delta \Phi^{(\nu)}_\alpha[\eta, \xi]=\!\!\! \int_0^t \!\!dt'\!\!\int_0^{t' }\!\!dt''\!
\left\{[\,\eta(t')\eta(t'')-\xi(t') \xi(t'')\, ]L^{(\nu)}_{\alpha,1} (t'-t'')\right.\nonumber\\
\fl &&\qquad\qquad \qquad\qquad\qquad\quad +\, \left.[\,\eta(t')\xi(t'') - \xi(t') \eta(t'')  \,] 
L^{(\nu)}_{\alpha,2} (t'-t'')\right\}     \, .   \label{eq:deltaphinu} 
\end{eqnarray}
The kernels are 
\begin{eqnarray}
\nonumber
L_\alpha(\tau)&=& L_\alpha'(\tau) + i L_\alpha''(\tau) = \frac{c_\alpha^2 q_0^2}{2m_\alpha\omega_\alpha}
 [\coth \frac{\beta \omega_\alpha}{2} \cos (\omega_\alpha\tau) - i\,  \sin(\omega_\alpha\tau) ]\, ,    
\\   \label{eq:kernels_l}
L^{(\nu)}_{\alpha,1} (\tau) &=&   \frac{c_\alpha^2 q_0^2}{2m_\alpha\omega_\alpha}
\frac{\sin(\nu\omega_\alpha/2) \sinh[(\beta+i\,\nu)\omega_\alpha/2]}{ \sinh(\beta\omega_\alpha/2)}\,\cos(\omega_\alpha\tau) \, ,     \\
 L^{(\nu )}_{\alpha,2} (\tau) &=&
\frac{c_\alpha^2 q_0^2}{2m_\alpha\omega_\alpha}
\frac{\sin(\nu\omega_\alpha/2) \cosh[(\beta+i\,\nu)\omega_\alpha/2]}{ \sinh(\beta\omega_\alpha/2)}\,\sin(\omega_\alpha\tau) \, .   \nonumber
\end{eqnarray}
\section*{Acknowledgements}
We acknowledge the support of the MIUR-FIRB2012 - Project HybridNanoDev (Grant  No.RBFR1236VV), EU FP7/2007-2013 under REA grant agreement no 630925 -- COHEAT, MIUR-FIRB2013 -- Project Coca (Grant
No.~RBFR1379UX), and the COST Action MP1209. U. W. acknowleges support from the Deutsche Forschungsgemeinschaft through SFB/TRR21.

\section*{References}
\providecommand{\newblock}{}


\begin{thebibliography}{10}
\expandafter\ifx\csname url\endcsname\relax
  \def\url#1{{\tt #1}}\fi
\expandafter\ifx\csname urlprefix\endcsname\relax\def\urlprefix{URL }\fi
\providecommand{\eprint}[2][]{\url{#2}}

\bibitem{esposito2009nonequilibrium}
Esposito M, Harbola U and Mukamel S 2009 {\it Rev. Mod. Phys.} {\bf
  81} 1665

\bibitem{esposito2009erratum}
Esposito M, Harbola U and Mukamel S 2014 {\it Rev. Mod. Phys.} {\bf 86}
  1125

\bibitem{campisi2011colloquium}
Campisi M, H{\"a}nggi P and Talkner P 2011 {\it Rev. Mod. Phys.}
  {\bf 83} 771

\bibitem{campisi2011erratum}
Campisi M, H{\"a}nggi P and Talkner P 2011 {\it Rev. Mod. Phys.}
  {\bf 83} 1653

\bibitem{levy2012quantumPRE}
Levy A, Alicki R and Kosloff R 2012 {\it Phys. Rev. E} {\bf 85} 061126

\bibitem{Kosloff-Levy2014}
Kosloff R and Levy A 2014 {\it Ann. Phys. Chem.} {\bf 65}
  365 
  
\bibitem{levy2012quantum}
Levy A and Kosloff R 2012 {\it Phys. Rev. Lett.} {\bf 108} 070604

\bibitem{campisi2009fluctuation}
Campisi M, Talkner P and H{\"a}nggi P 2009 {\it Phys. Rev. Lett.} {\bf
  102} 210401

\bibitem{kurchan2000quantum}
Kurchan J 2000 {\it ArXiv:0007360}

\bibitem{tasaki2000jarzynski}
Tasaki H 2000 {\it ArXiv:0009244}

\bibitem{talkner2007fluctuation}
Talkner P, Lutz E and H{\"a}nggi P 2007 {\it Phys. Rev. E} {\bf 75}
  050102

\bibitem{dorner2013extracting}
Dorner R, Clark S, Heaney L, Fazio R, Goold J and Vedral V 2013 {\it Phys. Rev. Lett.} {\bf 110} 230601

\bibitem{mazzola2013measuring}
Mazzola L, De~Chiara G and Paternostro M 2013 {\it Phys. Rev. Lett.}
  {\bf 110} 230602

\bibitem{batalho2014}
Batalh\~ao T~B, Souza A~M, Mazzola L, Auccaise R, Sarthour R~S, Oliveira I~S,
  Goold J, De~Chiara G, Paternostro M and Serra R~M 2014 {\it Phys. Rev. Lett.} {\bf 113} 140601

\bibitem{solinas2013work}
Solinas P, Averin D~V and Pekola J~P 2013 {\it Phys. Rev. B} {\bf 87}
  060508

\bibitem{watanabe2014generalized}
Watanabe G, Venkatesh B~P and Talkner P 2014 {\it Phys. Rev. E} {\bf 89}
  052116

\bibitem{pekola2013calorimetric}
Pekola J, Solinas P, Shnirman A and Averin D 2013 {\it New J. Phys.} {\bf 15} 115006

\bibitem{gasparinetti2014fast}
Gasparinetti S, Viisanen K, Saira O~P, Faivre T, Arzeo M, Meschke M and Pekola
  J 2014 {\it ArXiv:1405.7568}

\bibitem{gasparinetti2014heat}
Gasparinetti S, Solinas P, Braggio A and Sassetti M 2014 {\it New J. Phys.} {\bf 16} 115001

\bibitem{silaev2014lindblad}
Silaev M, Heikkil{\"a} T~T and Virtanen P 2014 {\it Phys. Rev. E} {\bf
  90} 022103

\bibitem{wollfarth2014distribution}
Wollfarth P, Shnirman A and Utsumi Y 2014 {\it Phys. Rev. B} {\bf 90}
  165411

\bibitem{weiss1999quantum}
Weiss U 2012 {\it Quantum dissipative systems}, 4th edition,  (World Scientific, Singapore).

\bibitem{ingold2002path}
Ingold G~L 2002 Path integrals and their application to dissipative quantum
  systems {\it Coherent Evolution in Noisy Environments} (Springer) pp 1--53

\bibitem{orth2010universality}
Orth P~P, Imambekov A and Le~Hur K 2010 {\it Phys. Rev. A} {\bf 82}
  032118

\bibitem{orth2013nonperturbative}
Orth P~P, Imambekov A and Le~Hur K 2013 {\it Phys. Rev. B} {\bf 87}
  014305

\bibitem{henriet2014quantum}
Henriet L, Ristivojevic Z, Orth P~P and Le Hur K~L 2014 {\it Phys. Rev. A} {\bf 90} 023820

\bibitem{bulla2003numerical}
Bulla R, Tong N~H and Vojta M 2003 {\it Phys. Rev. Lett.} {\bf 91}
  170601

\bibitem{egger1994low}
Egger R and Mak C 1994 {\it Phys. Rev. B} {\bf 50} 15210

\bibitem{sassetti1990universality}
Sassetti M and Weiss U 1990 {\it Phys. Rev. Lett.} {\bf 65} 2262

\bibitem{grifoni1996exact}
Grifoni M, Sassetti M and Weiss U 1996 {\it Phys. Rev. E} {\bf 53} R2033

\bibitem{keil2001real}
Keil M and Schoeller H 2001 {\it Phys. Rev. B} {\bf 63} 180302

\bibitem{grifoni1995cooperative}
Grifoni M, Sassetti M, H{\"a}nggi P and Weiss U 1995 {\it Phys. Rev. E}
  {\bf 52} 3596

\bibitem{kopp2007universal}
Kopp A and Le~Hur K 2007 {\it Phys. Rev. Lett.} {\bf 98} 220401

\bibitem{grifoni1998driven}
Grifoni M and H{\"a}nggi P 1998 {\it Phys. Rep.} {\bf 304} 229

\bibitem{kehrein1996spin}
Kehrein S~K and Mielke A 1996 {\it Phys. Lett. A} {\bf 219} 313

\bibitem{grifoni1993nonlinear}
Grifoni M, Sassetti M, Stockburger J and Weiss U 1993 {\it Phys. Rev. E}
  {\bf 48} 3497

\bibitem{grifoni1997dissipative}
Grifoni M, Hartmann L and H{\"a}nggi P 1997 {\it Chem. Phys.} {\bf 217}
  167

\bibitem{hartmann2000driven}
Hartmann L, Goychuk I, Grifoni M and H{\"a}nggi P 2000 {\it Phys. Rev. E} {\bf 61} R4687

\bibitem{caldeira1983path}
Caldeira A~O and Leggett A~J 1983 {\it Physica A} {\bf 121} 587

\bibitem{leggett1987dynamics}
Leggett A~J, Chakravarty S, Dorsey A, Fisher M~P, Garg A and Zwerger W 1987
 {\it Rev. Mod. Phys.} {\bf 59} 1

\bibitem{gorlich1989low}
G{\"o}rlich R, Sassetti M and Weiss U 1989 {\it Eur. Phys. Lett.}
  {\bf 10} 507

\bibitem{feynman1963theory}
Feynman R~P and Vernon~Jr F 1963 {\it Ann. Phys.} {\bf 24} 118

\bibitem{grifoni1999dissipation}
Grifoni M, Paladino E and Weiss U 1999 {\it Eur. Phys. J. B} {\bf 10} 719

\bibitem{grifoni1997coherences}
Grifoni M, Winterstetter M and Weiss U 1997 {\it Phys. Rev. E} {\bf 56}
  334

\end{thebibliography}
\end{document}